\documentclass[11pt,a4paper,onecolumn,reqno]{amsart}
\usepackage[a4paper, margin=2.3cm, bmargin=3cm]{geometry}

\usepackage{tabulary}
\usepackage{booktabs}
\usepackage{graphicx,stfloats}
\usepackage{epstopdf}
\usepackage{color}
\usepackage{amsmath}
\usepackage{amsaddr}
\usepackage{bm, soul}
\usepackage{mathtools}
\usepackage[colorinlistoftodos,prependcaption,textsize=tiny]{todonotes}
\usepackage{braket}
\usepackage[hyphens]{url}
\usepackage{hyperref}
\usepackage{makecell}
\usepackage{array}
\usepackage{placeins}
\usepackage[square,comma,sort&compress, numbers]{natbib}

\usepackage[version=3]{mhchem} \usepackage{siunitx}
 \usepackage{multirow}

\usepackage{etoolbox}
\patchcmd{\pprintMaketitle}
  {\hrule\vskip12pt}
 {\hrule\vskip12pt\ifvoid\extrainfobox\else\unvbox\extrainfobox\par\vskip12pt\fi}
 {}{}

\newsavebox\extrainfobox

\usepackage{caption}
\usepackage[utf8]{inputenc}
\usepackage{amssymb}
\usepackage{amsmath}
\usepackage{hyperref}
\usepackage{ulem}
\usepackage[export]{adjustbox}
\usepackage{xcolor}
\usepackage{nicefrac}
\usepackage{tabu,multirow}
\usepackage{tabularx}
\def\hlinewd#1{%
\noalign{\ifnum0=`}\fi\hrule \@height #1 %
\futurelet\reserved@a\@xhline}
\usepackage{graphicx}
\usepackage{amssymb}

\usepackage{lineno}
\usepackage{color}
\usepackage{float}
\usepackage{epstopdf}
\usepackage{siunitx}
\usepackage{hyphenat}
\usepackage{todonotes}
\usepackage{subcaption}

\DeclareSIUnit\volper{vol\%}
\DeclareSIUnit\massper{\%_{m}}
\usepackage{array}
\DeclareMathAlphabet{\mathpzc}{OT1}{pzc}{m}{it}
\newcolumntype{P}[1]{>{\centering\arraybackslash}p{#1}}
\hypersetup{
colorlinks=true,
linkcolor=blue,
citecolor=blue,
citebordercolor={0 1 0}
}

\title{Techno-economic assessment of long-distance supply chains of energy carriers: Comparing hydrogen and iron for carbon-free electricity generation}

\author[TTD]{Jannik Neumann$^1$, Rodolfo Cavaliere da Rocha$^2$, Paulo Debiagi$^2$, Arne Scholtissek$^{2,*}$, Frank Dammel$^1$, Peter Stephan$^1$, Christian Hasse$^2$}
\email{scholtissek@stfs.tu-darmstadt.de} 

\address[]{$^1$ Technical University of Darmstadt, Department of Mechanical Engineering, Institute for Technical Thermodynamics, Alarich-Weiss-Straße 10,  64287 Darmstadt, Germany\\
$^2$Technical University of Darmstadt, Department of Mechanical Engineering, Simulation of reactive Thermo-Fluid Systems, Otto-Berndt-Str. 2, 64287 Darmstadt, Germany}

\begin{document}
\pagestyle{plain}
\maketitle

\begin{abstract}
\label{sec:Abstract}
The effective usage of renewable energy sources requires ways of storage and delivery to balance energy demand and availability divergences. Carbon-free chemical energy carriers are proposed solutions, converting clean electricity into stable media for storage and long-distance energy trade. Among them, hydrogen (\ce{H2}) is noteworthy, being the subject of significant investment and research. Metal fuels, such as iron (\ce{Fe}), represent another promising solution for a clean energy supply, but establishing an interconnected ecosystem still requires considerable research and development. This work proposes a model to assess the supply chain characteristics of hydrogen and iron as clean, carbon-free energy carriers and then examines case studies of possible trade routes between the potential energy exporters Morocco, Saudi Arabia, and Australia and the energy importers Germany and Japan. The work comprehends the assessment of economic (levelized cost of electricity - LCOE), energetic (thermodynamic efficiency) and environmental (\ce{CO2} emissions) aspects, which are quantified by the comprehensive model accounting for the most critical processes in the supply chain. The assessment is complemented by sensitivity and uncertainty analyses to identify the main drivers for energy costs. Iron is shown to be lower-cost and more efficient to transport in longer routes and for long-term storage, but potentially more expensive and less efficient than \ce{H2} to produce and convert. Uncertainties related to the supply chain specifications and the sensitivity to the used variables indicate that the path to viable energy carriers fundamentally depends on efficient synthesis, conversion, storage, and transport. A break-even analysis demonstrated that clean energy carriers could be competitive with conventional energy carriers at low renewable energy prices, while carbon taxes might be needed to level the playing field. Thereby, green iron shows potential to become an important energy carrier for long-distance trade in a globalized clean energy market.\\
\end{abstract}


\clearpage

\section{Introduction}
\label{sec:Introduction}
Fossil fuels were still the source of over \SI{80}{\%} of the global energy supply in 2021 \cite{IEAwebsite,BP2021}, bringing annual \ce{CO2} emissions to the historical peak of 36.3\,Gt. The most recent assessment by the Intergovernmental Panel on Climate Change (IPCC) predicts that, without a more intense effort in decarbonization, drastic climate changes with catastrophic consequences for the environment and human society are inevitable \cite{AR6full2021}. During the 2021 United Nations Climate Change Conference (COP26) \cite{COP26}, an agreement on phasing down unabated coal power was reached, accelerating the decommissioning of coal power plants even before sustainable replacements are available.

Besides the climate crisis, the globalized world is facing a surge in energy demand, which adds to several complications in the transport of commodities worldwide. Moreover, the recent war against Ukraine raises additional uncertainties on the future availability, prices, and security of supply routes for fossil fuels, including coal, oil and natural gas (NG), which are directly influenced by sanctions and retaliatory movements enacted by governments \cite{forbes,wef2022}.

The European Union as a whole, and Germany in particular, heavily rely on energy imports. Therefore, the need for diversification of the European energy mix is increasingly urgent. This process shall involve phasing down fossil fuel consumption and accelerating the transition to locally available (i.e. domestic) renewable energy. Nevertheless, this change comes with several challenges. Renewable sources such as wind and solar radiation are, in general, converted directly to electricity, which cannot replace traditional fuels in several applications. Further, as the share of renewables increases in the energy mix, maintaining grid stability also becomes an issue, currently handled by backup thermal power plants running on coal, NG or petroleum products. Finally, several countries exhibit limited renewable energy potential, preventing their complete energetic self-sufficiency in a decarbonized future and therefore rely on energy imports for years to come \cite{SRU2021}. Efficiently converting renewable electricity into green energy carriers (ECs) is a proposed solution. In this stable form, their energy content can be transported, traded, stored and utilized when and where power is required, being considered by some \cite{bergthorson2018} as the key to unlocking a sustainable, cost-effective and safeguarded energy supply.

Electrical energy can be stored by electric, (electro-)chemical, mechanical and thermal means. Due to their high volumetric energy density, stability and versatility, chemical ECs are suggested to enable long-distance energy trade and clean remote power generation and also to serve as alternative fuel sources for heavy-duty machinery and vehicles~\cite{bergthorson2018}. Among several proposed candidates, carbon-free ECs such as hydrogen~\cite{thefutureofhydrogen}, ammonia~\cite{nh3forpower,valera-medina2021}, and metals~\cite{BERGTHORSON2015368} are considered some of the most promising, and have been the focus of intense research and investment recently. The first two are well-known for their potential and applications, while the possibilities of metals such as iron or aluminium as ECs are still to be explored, demanding more research and development~\cite{julien2017,hazenberg2021,debiagi2022,ning2022}.

Due to its versatility as EC and chemical precursor, hydrogen is gaining unprecedented momentum. In the context of the hydrogen economy, converting electricity into \ce{H2} is touted as a way of storing energy from renewable sources~\cite{thefutureofhydrogen}. However, due to its very high reactivity, diffusion characteristics, and low volumetric energy density, it is challenging to transport \ce{H2} over long distances or to store it for long periods~\cite{hampp2022,JOHNSTON2022}. In this context, metal-based cycles can offer advantages by providing low-cost, safe and energy-efficient storage and transport of the EC. Iron, for instance, can be produced with clean hydrogen--therefore generating no direct \ce{CO2} emission--and can either be used in its synthesis or directly applied in burners, such as those of existing coal power plants for electricity generation, through retrofitting ~\cite{bergthorson2018,thefutureofhydrogen,ISTR2020,Janicka2022}. By applying this concept, time and money can be spared in developing new systems and infrastructure construction, becoming an interesting technology for countries that still rely on coal due to the absence of ready-to-use alternatives~\cite{debiagi2022}.

\begin{figure}[!h]
		\begin{center}
        \includegraphics[width=\textwidth]{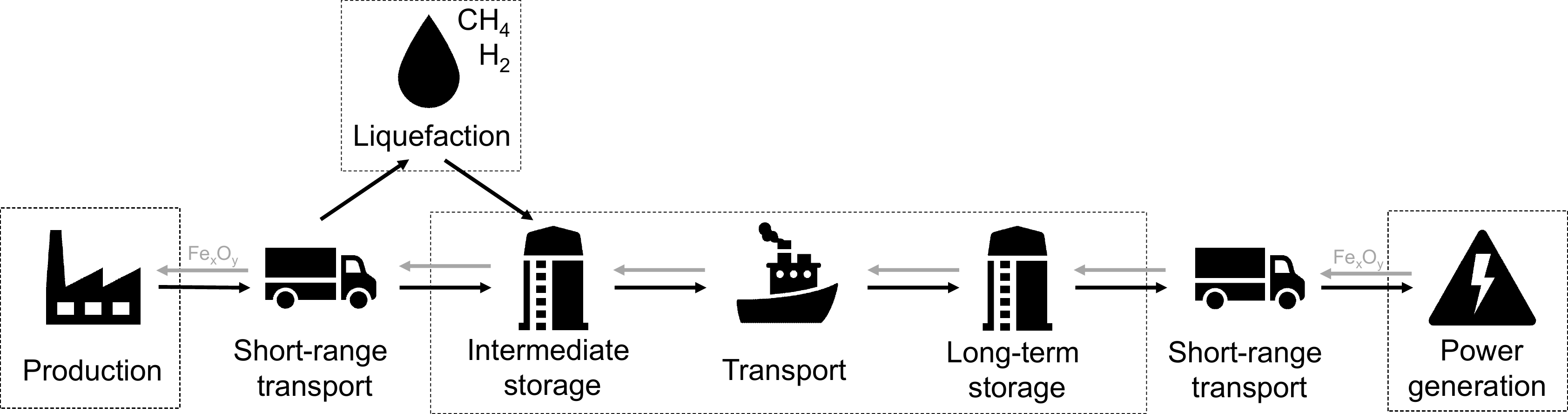}
        \end{center}
        \caption{Schematic for the supply chain of an EC from production to utilization. Dashed boxes mark the processes considered in the current work.}
		\label{fig:TransportDiagram}
\end{figure}

A complete feasibility study of an EC must consider the aspects of each process involved in the supply chain, which is schematically shown in Fig.~\ref{fig:TransportDiagram}. While numerous studies on hydrogen and hydrogen-based ECs exist~\cite{JOHNSTON2022, Heuser.2019, Crandall.2022, Hank.2020, IRENA.2022,  Wijayanta.2019, Ishimoto.2020, Al-Breiki.2020}, similar studies for iron as EC are still very scarce. Kuhn et al.~\cite{Kuhn.2022} calculate a cycle efficiency of \SI{27}{\%} for iron as EC not considering transport. Debiagi et al.~\cite{debiagi2022} estimate a round-trip efficiency of \SI{26}{\%}-\SI{31}{\%} including transport by ship from Casablanca to Rotterdam. Dirven et al.~\cite{DIRVEN201852} estimates a cycle efficiency of \SI{15}{\%}-\SI{30}{\%}, including power generation, transport and regeneration of metal fuels. The first and only available techno-economic evaluations for iron as EC are published by Metalot, a dutch non-profit network organisation~\cite{Hajonides.2022, Metalot.2022}. Their analyses relate to high-temperature heat supply, and they conclude that the cost of iron is in the range of hydrogen and superior to ammonia as EC for high-temperature heat supply.

This work presents an assessment and comparison of carbon-neutral hydrogen (green \ce{H2}) and iron (green \ce{Fe}) as chemical ECs for long-distance energy trade, considering aspects of the combined supply chain, including synthesis, transport, storage and utilization in electricity production (c.f.~Fig.~\ref{fig:TransportDiagram}), and the impact of each step in the total costs, \ce{CO2} emissions and energy efficiency. To the best of the authors' knowledge, this work presents the first comprehensive techno-economic analysis and comparison between iron and other ECs for electricity generation. As potential clean EC importers, Germany and Japan are taken as case studies, considering aspects of their energy mix, port infrastructure and retrofitting possibilities. As potential EC exporters, Morocco, Saudi Arabia and Australia are taken considering their energy mix, industry, and pledged future policies. The proposed comprehensive formulation are employed for estimations of costs, emissions, and energy efficiencies. Finally, a sensitivity analysis to evaluate the effect of critical variables, an uncertainty analysis to determine the extent of variation between available data, and a break-even analysis to determine the point of competitiveness between green and conventional ECs are presented.

\section{Properties, production and power generation methods of selected energy carriers}

Long-distance energy trade and efficient energy storage are essential to safeguard the supply chain and grid stability when decarbonizing the energy mix. In order to select the most suitable carriers, it is necessary to assess aspects of their production and utilization, costs and associated \ce{CO2} emissions, and their physical-chemical characteristics. Thus, this section brings relevant information for conventional fuels, hydrogen and iron.

\subsection{Relevant properties of selected chemical energy carriers}
\label{sec:fuel_properties}

\begin{figure}[!h]
		\begin{center}
        \includegraphics[width=0.5\textwidth]{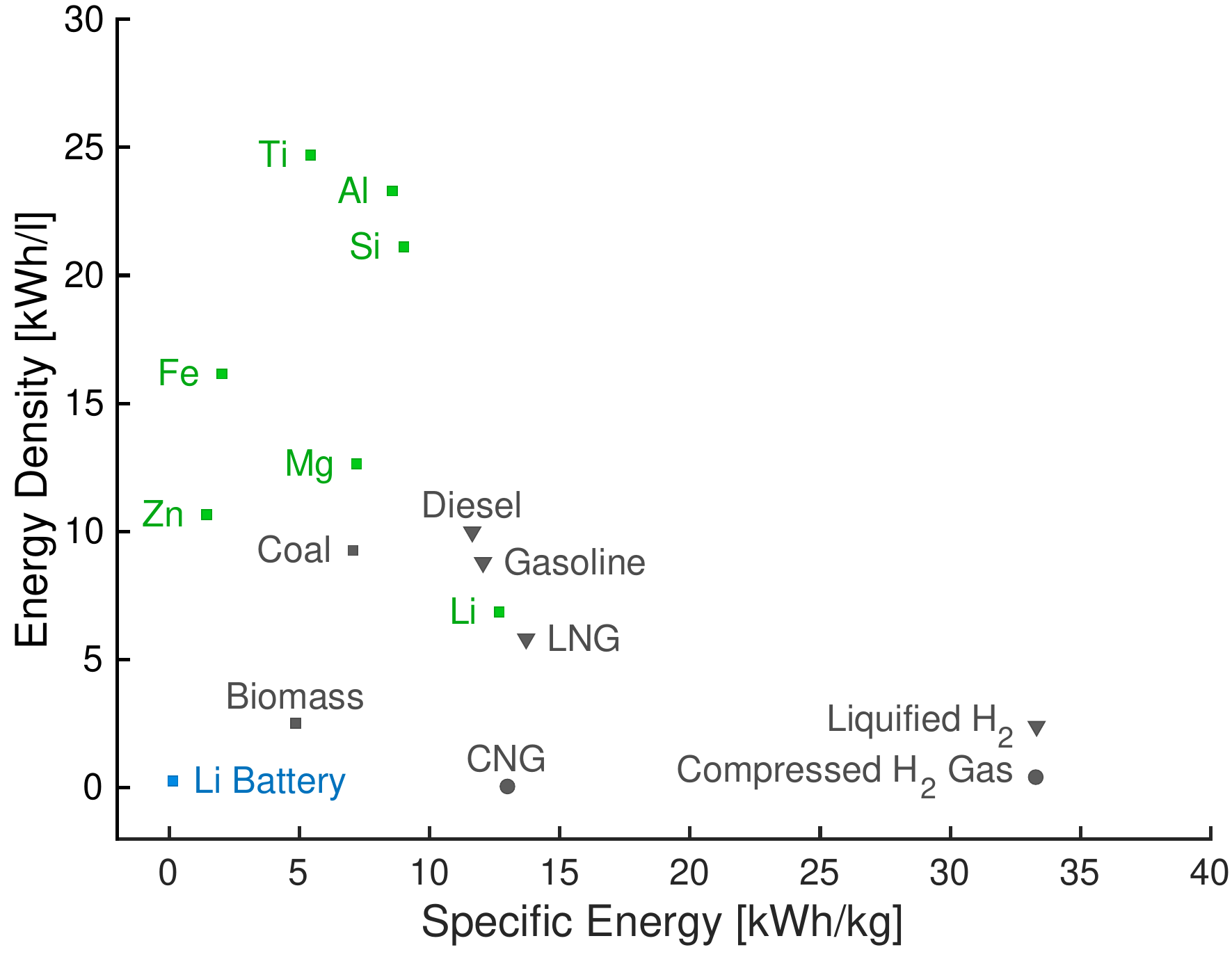}
        \end{center}
        \caption{\textbf{Volumetric and gravimetric energy densities of selected chemical ECs}.  Representative hydrocarbon fuels are characterized by a balance between the two properties. \ce{H2} has a high gravimetric energy density but a low volumetric density, even when compressed or liquefied. In contrast, metals have a very high volumetric energy density with a low to medium gravimetric energy density.}
		\label{fig:EnergyDensity}
\end{figure}

The energy density is an essential feature of ECs. Substances that present high volumetric and/or gravimetric energy densities, commonly achieved when in liquid or solid phases, are desirable, requiring smaller storage and cargo units for a certain amount of internal energy, and therefore fewer trips, saving in fuel, equipment and other costs. The energy densities of selected substances are shown in Fig.~\ref{fig:EnergyDensity}.

Significant properties of selected chemical ECs are reported in Table~\ref{tab:EC_properties}. Iron is a very dense material, has lower gravimetric energy density when compared to coal, but its volumetric energy density is significantly higher, reducing the required transport and storage unit volumes. NG and hydrogen reach volumetric energy densities in the same order of magnitude as liquid and solid ECs only when compressed or liquefied for transport.

\begin{table}[!h]\scriptsize
\caption{Properties of selected chemical ECs~\cite{nist,LETT2004411}.}\label{tab:EC_properties}
\centering
\begin{tabular}{l c c c c}
    \hline
         & Density                       & Boiling point     & Grav. energy density               & Vol. energy density  \\
                        & [kg $\cdot \ \mathrm{m}^{-3}$]                 & [\textdegree C]   & [kWh $\cdot \ \mathrm{kg}^{-1}$]    & [kWh l$^{-1}$] \\ 
    \hline
    Coal            & 1200--1600                    & n.a.              & 7--10  & 8.3--15.8 \\
    Iron (\ce{Fe})      & 7870                          & n.a.              & 2.1 & 16.1 \\ 
    Nat. Gas (\ce{CH4}) & 0.66$^1$/410$^2$              & -161.6            & 13.9 & 0.009$^1$/ 5.7$^2$ \\ 
    Hydrogen (\ce{H2})  & 0.08$^1$/30$^3$/57$^4$/73.5$^5$ & -259.9          & 33.3& 0.0025$^1$/ 1$^3$/ 1.89$^4$/ 2.61$^2$ \\ 
    \hline
    \multicolumn{5}{l}{$^1$STP, $^2$liquid, $^3$350 bar, $^4$700 bar}
\end{tabular}
\end{table}

\subsection{Conventional fuels: oil, gas and coal}
\label{sec:background}

Coal, petroleum and NG are the primary global sources of energy for transport, power generation and heating applications, and are also the main contributors to \ce{CO2} emissions and global warming~\cite{WEO2021,AR6full2021}. In 2019, oil, NG and coal combined supplied 1.36\,EWh of energy and generated 33\,Gt of \ce{CO2}~\cite{WEO2021}. In the power sector alone, the share of coal reached \SI{36.6}{\%} in 2021 (10\,PWh), while the share of NG has risen to \SI{22.1}{\%} (6.1\,PWh), and that of oil and other fossil fuels, to \SI{3}{\%} (0.81\,PWh). 11.8\,Gt of \ce{CO2} were emitted by the electricity sector alone, of which 8.23\,Gt came from coal, 2.97\,Gt from NG, and 0.57\,Gt from oil and other fossils~\cite{emberNew}. A clear pattern of fluctuation in trade prices for conventional fuels is evident from the recent data shown in Table~\ref{table:FF_production}, which varies based on time and the specific trading location. Additionally, increasingly strict policies on \ce{CO2} emissions, including carbon taxes~\cite{EU.2022b, EU.2022,carbontax}, is intended to discourage fossil-fuel use. Current EU carbon prices (end of 2022) are in the order of 90\,EUR$\cdot\mathrm{t}_{\ce{CO2}}^{-1}$~\cite{Ember.2022}, but the IEA predicts them to rise to 130\,USD$\cdot\mathrm{t}_{\ce{CO2}}^{-1}$ by 2030 and to 250\,USD$\cdot\mathrm{t}_{\ce{CO2}}^{-1}$ by 2050 across all advanced economies \cite{IEA.2021b}. Coupled with the drastic reduction of renewable energy costs~\cite{thefutureofhydrogen} and the fact that only a few countries control fossil fuel markets, renewables are soon expected to become more economically attractive~\cite{thefutureofhydrogen,IRENA2020}, overcoming a fundamental challenge of the energy transition.

\begin{table}[!h]\scriptsize
\caption{\textbf{Recent and historic prices of conventional fossil fuels.} The prices shown represent the one-month futures of the Dutch TTF Natural Gas~\cite{RotterdamNG.23} and Henry Hub~\cite{NaturalGasHenry.23} for NG as well as the API2 Rotterdam Coal Futures~\cite{RotterdamCoal.23} and Newcastle Coal Futures~\cite{NewCastleCoal.23} for coal. Coal properties are taken from the average of the values presented in Table~\ref{tab:EC_properties}.}
\centering
\begin{tabular}{l  c  c c c  }
    \hline
     Fuel & \multicolumn{2}{c}{Coal} & \multicolumn{2}{c}{Natural gas}  \\
    \hline
     &  \multicolumn{4}{c}{Price [$\mathrm{USD} \cdot  \mathrm{MWh}^{-1}$] }     \\
    Date & Rotterdam &  Newcastle &Rotterdam & Henry Hub\\
    \cmidrule(lr){2-2}\cmidrule(lr){3-3}\cmidrule(lr){4-4}\cmidrule(lr){5-5}
    01.2023 & 15.58 & 29.71& 54.62 & 9.16 \\
    06.2022 & 43.67 &45.86 & 137.63 & 18.51\\
    01.2022 & 21.04 & 22.09& 80.64 &16.63\\
    06.2021 & 14.25 & 14.97 & 32.97 & 12.45\\
    01.2021 & 8.00  &  8.40 & 18.88 & 8.75\\
    \hline
    \end{tabular}
\label{table:FF_production}
\end{table}

\subsection{Hydrogen}
Hydrogen (\ce{H2}) is touted as an ideal EC due to its optimal physical-chemical properties, low toxicity, versatility and simplicity of production. As a zero-carbon fuel, its combustion does not produce any \ce{CO2}. It can be employed in numerous power production systems, to provide heat to industries that still rely on fossil fuels, and also as a reducing agent in chemical processes. For its potential, hydrogen has become the focus of extensive research~\cite{thefutureofhydrogen} and investments~\cite{SRU2021,h2europe}. The Net Zero Emissions Scenario from the IEA~\cite{IEANetZero} predicts a hydrogen demand of 530\,Mt by 2050 driven by the transport, industry and electricity sectors. Within the power sector, the use of hydrogen is expected to increase significantly (102\,Mt in 2050 \cite{IEANetZero}), as it can help balance the increasing generation from intermittent solar and wind power by storing seasonal energy for future use.

Presently, 90\,Mt of pure hydrogen are produced annually, primarily used in oil refining (\SI{44}{\%}) and for chemicals production (\SI{45}{\%}, of which $\frac{3}{4}$ for ammonia and $\frac{1}{4}$ for methanol)~\cite{IEA.2021}. Another 5\,Mt of non-pure hydrogen are used in steelmaking, and marginal amounts for heat and power~\cite{IEA.2021}. Hydrogen can be converted in fuel cells~\cite{FCfactsheet2015} and gas turbines~\cite{ETN.2020} for power production. Hydrogen-enriched NG combustors are already commercially available~\cite{ETN.2020,Mitsubishi.2018}, while pure hydrogen-burning commercial turbines are under development \cite{GE.2022,Mitsubishi.2018}.

Green-\ce{H2} production is becoming highly efficient. However, it still lacks scale, being only \SI{0.7}{\%} of the total pure hydrogen produced worldwide~\cite{thefutureofhydrogen, IEA.2021}. Governments and industry are pushing the increase in electrolyzer capacity, with existing plans in Germany~\cite{SRU2021}, the European Union~\cite{h2europe}, Australia~\cite{augovmnt2022} and Japan ~\cite{japan.2022,KOBAYASHI2019109}. Based on current and future initiatives, green \ce{H2} production could reach 8\,Mt by 2030, which is 4\,Mt less than the required hydrogen production capacity in the IEA's Announced Pledges Scenario and 72\,Mt less compared to the Net Zero Emissions Scenario by 2030 \cite{IEA.2021b}. The main issues of green-\ce{H2} are still the high cost of infrastructure and the dependency on electricity prices \cite{thefutureofhydrogen,h2europe}. 

\subsection{Iron}
\begin{figure}[!h]
		\begin{center}
        \includegraphics[width=0.5\textwidth]{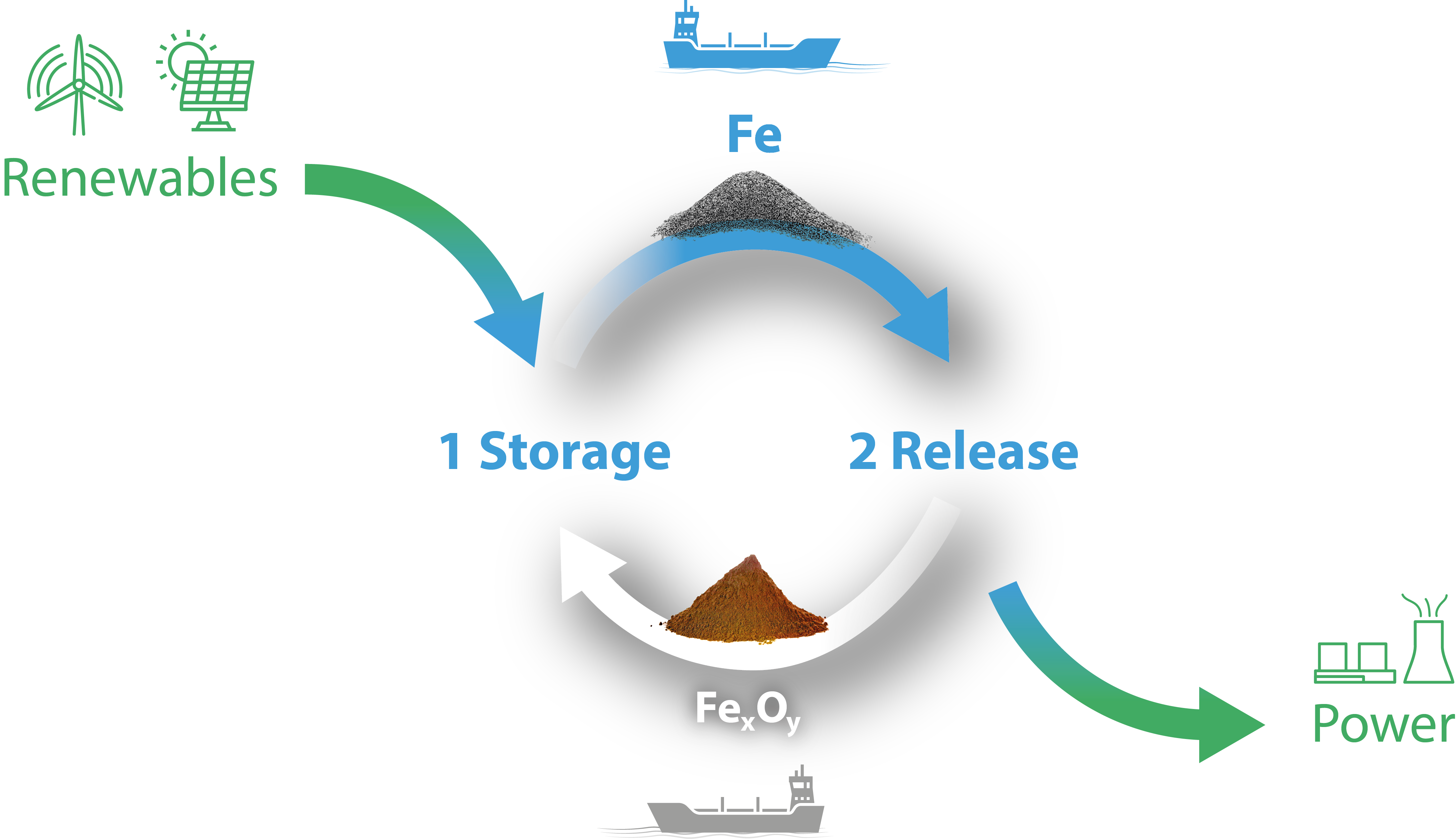}
        \end{center}
        \caption{Schematic of an iron reduction-oxidation cycle for energy storage, transport and use~\cite{clean_circle}. Iron and iron oxides are used in a reduction-oxidation cycle as carbon-free carriers of renewable energy. Left: renewable energy is used to reduce iron oxides via electrochemical (i.e. electrolysis) or thermochemical processes (i.e. reacting oxides with green hydrogen), storing the energy. Right: iron is used as a fuel in a combustor to release heat, e.g. in a power plant, to generate electricity, similarly to the traditional combustion of solid fuels. Solid iron oxides (Fe$_x$O$_y$) can be collected and transported to the reduction facilities, closing the cycle.}
		\label{fig:cleancircles}
\end{figure}

The idea of using iron (\ce{Fe}) as a green EC has been proposed recently~\cite{bergthorson2018,BERGTHORSON2015368,julien2017}, despite having been idealized as a fuel for more than a century~\cite{DIRVEN201852}. Iron is chemically stable, has low toxicity, abundance, and a very high volumetric energy density, and is relatively low-cost to produce in processes currently evolving towards a zero-carbon footprint~\cite{ArcelorMittal,HYBRIT,thyssenkrupp}. These characteristics are ideal for an EC. Furthermore, iron has burning characteristics similar to those of coal~\cite{BERGTHORSON2015368,debiagi2022}, reaching very high temperatures when burned in clouds of microparticles, generating zero \ce{CO2} emissions, potentially enabling its use in retrofitted coal power plants~\cite{bergthorson2018,debiagi2022} which allows the reuse of soon-to-be redundant infrastructure and short implementation times. The stable products of its combustion are solid iron oxides, which are non-toxic and can be collected and recycled back into metallic iron form. The recycling (reduction) can be carried out cleanly and efficiently, directly using renewable electricity or green-\ce{H2}. In that case, it can become the core of a clean redox cycle, showing high potential for long-term storage and transport of renewable energy, see Fig.~\ref{fig:cleancircles}.

The iron production processes in the steel industry involve the reduction of $\mathrm{Fe}_x\mathrm{O}_y$ oxides present in iron ore, being the blast-furnace--basic oxygen furnace (BF-BOF) the most common~\cite{ISTR2020}, in which the oxides are reduced using coke (i.e. heated coal in the absence of oxygen), with the resulting material being later reacted with limestone to release excess carbon in the form of \ce{CO2}. This process is highly endothermic and traditionally uses fossil fuels to generate heat in both steps. However, an emerging technology is the direct reduction of iron (DRI) using methane, syngas or hydrogen to reduce the oxides in iron ore pellets, followed by melting and alloying in an electric arc furnace (EAF)~\cite{ISTR2020}.

Green iron production requires decarbonizing both direct and indirect \ce{CO2} emissions. One solution is to perform DRI with \ce{H2} and heat provided by renewable sources~\cite{ISTR2020,MIGNARD20075039,LORENTE20095554,Vogl.2018}, an alternative being pursued by important industrial players~\cite{ArcelorMittal,HYBRIT,thyssenkrupp}. The overall energy efficiency of the process can achieve 60\% or more, depending strongly on the electrolyzer efficiency that generates the hydrogen \cite{Vogl.2018, Bhaskar.2020}.

\section{Transport and storage of energy carriers}
\label{sec:transportation}

From production to utilization of the ECs, a long supply chain includes preparation, loading, unloading, and delivery to the final application, which is schematically shown in Fig.~\ref{fig:TransportDiagram}. Between these steps, intermediate storage is important, balancing supply and demand. For gaseous ECs, liquefaction is necessary for effective storage and transport. All those steps increase costs, add to the energy demand, and contribute to \ce{CO2} emissions. This section presents conventional and alternative modes of long-distance transport and storage, along with the perspectives in terms of technology and infrastructure.

\subsection{Preparation: liquefaction of gaseous fuels}
The liquefaction of NG is operated today on a global scale with an existing liquefaction capacity of 1.26\,million metric tons per day in 2021 \cite{IGU.22}. Refrigeration cycles are used to liquefy NG, in which the NG or a refrigerant cools down through compression, heat dissipation and expansion \cite{Zhang.2020}. 

The total global capacity of hydrogen liquefiers is 350\,metric tons per day, with the largest liquefier having a capacity of 32\,metric tons per day~\cite{Ghafri.22}. Existing hydrogen liquefiers require \si{30}--\SI{60}{\%} of the chemical bound energy for the liquefaction~\cite{Krasae.2010, Ghafri.22, Ohlig.2014, Majid.2022}. It is predicted that the specific energy demand can be further reduced to as low as \SI{13}{\%}~\cite{Ghafri.22,Majid.2022} with new concepts and larger plants (lower specific insulation losses). Furthermore, due to learning curves and economies of scale, the specific capital costs (CC) might be reduced significantly. However, these improvements refer to concepts with low technology readiness levels, and it remains to be shown if these numbers can be obtained with commercially available \ce{H2}-liquefiers in a few years. Table~\ref{table:liquefaction} shows the ranges of critical liquefaction-related values, while a compilation of literature values is reported in Table~S~3.

\subsection{Long-distance maritime transport: vessel types, capacities and fuel consumption} \label{sec:maritimeTransport}
Currently, long-distance maritime trade for solid ECs is feasible using bulk cargo ships, i.e. vessels that store their cargo freely in internal deposits. Conversely, gaseous fuels are transported as liquefied cargo via liquid bulk ships, requiring refrigeration to maintain temperatures below their condensation point throughout the trip. Coal and iron, for instance, can be transported in solid form at room temperature and atmospheric pressure, simplifying the loading and maintenance processes. A significant drawback, however, is that due to the reactivity of iron dust, powder iron cargo must be stored either in inert gas or sealed in appropriate containers, such as super bags. In terms of propulsion, most ships use heavy fuel oil (HFO) in their engines, while some LNG carriers use NG released from the boil-off to generate power.

No ship built to this day can transport hydrogen in volumes comparable to the cargo units of large LNG ships, the state-of-the-art in terms of liquid bulk vessels. The only currently available vessel is the Suiso Frontier, a diesel-propelled liquid hydrogen (L\ce{H2}) pilot ship, with a dead-weight tonnage (DWT) of 8000\,t and a cargo volume of 1250\,$\mathrm{m}^3$~\cite{Kawasaki.2019}. There, \ce{H2} is kept liquid at -253\,\textdegree C~\cite{bairdmaritime} through thermal insulation. An LNG ship with a similar DWT could carry considerably larger cargo, supposedly due to the space used for L\ce{H2} insulation. There are ongoing projects for future \ce{H2} vessels, capable of carrying 160000\,$\mathrm{m}^3$ of L\ce{H2}~\cite{Kawasaki.2022}--roughly 128 times the capacity of the Suiso Frontier. Another example is the C-Job Naval Architects plan of a new, 37500\,$\mathrm{m}^3$ fuel-cell-powered vessel, expected to be commissioned by 2027~\cite{C-Job.22}. Provaris Energy is at conceptual stage of development of a compressed (250\,bar) \ce{H2} carrier~\cite{provaris}. In terms of propulsion, there are already plans for hydrogen-fuelled internal combustion engines, which could directly use the stored cargo as fuel, making use of the boil-off effect~\cite{MAN}.

\subsection{Storage of energy carriers}
\label{sec:storage}

The viability of ECs also depends on intermediate and long-term storage. The requirements for the different ECs vary significantly, in some cases having a substantial effect on costs and overall efficiency. Coal is usually transported via trains or trucks from a coal mine to a harbour, where it is unloaded and piled~\cite{Ernst.2008}. Due to its low reactivity at ambient conditions and solid state, no special processing is required~\cite{Ernst.2008}. On the other hand, iron shows reactivity with air at ambient conditions, which would result in oxidation, consequently depleting the energy content and posing risks of ignition. Therefore, fine iron powders should be isolated from the ambient air, such as in sealed containers. Conversely, iron oxides produced from combustion processes are generally inert, requiring no special material handling. To be kept liquid, LNG and \ce{LH2} demand the maintenance of cryogenic temperatures, i.e, -162\,\textdegree C and -253\,\textdegree C, respectively, which lead to additional energy expenditures.

Global LNG storage capacity was nearly 71\,million\,$\mathrm{m}^3$ as of April 2022~\cite{IGU.22}. Despite great insulation efforts, some heat absorption from the environment cannot be circumvented due to the significant temperature difference between the tank content and the surroundings. The absorbed heat is balanced by the evaporation of some of the liquid (latent heat of vaporization), leading to boil-off gas (BOG). Additional energy is required (0.5--1.27\,$\mathrm{kWh} \cdot  \mathrm{kg}_{BOG}^{-1}$ ~\cite{Pospisil.2019,Air_Liquide.2016, Kim.2019}) to re-liquefy the inevitable BOG.

The global L\ce{H2} storage capacities are orders of magnitudes smaller, with the largest active L\ce{H2} tank having 3200 $\mathrm{m}^3$, used as a rocket fuel storage operated by NASA, with a boil-off rate (BOR) of 0.064 wt.\% per day \cite{Fesmire.2021}.

To suit the projected large-scale L\ce{H2} vessels (e.g. 160000 $\mathrm{m}^3$), the required tank scales for intermediate storage to enable international transport of L\ce{H2} by ships are two orders of magnitude larger and have yet to be demonstrated.
Similarly to LNG storage tanks, BOG cannot be entirely prevented, needing to be re-liquefied or flared off if no demand application can be coupled. Moreover, a recently built storage tank by NASA \cite{Fesmire.2021} features an integrated refrigeration system, leading to net zero boil-off \cite{Fesmire.2021, Swanger.21, Swanger.2022}, with a trade-off of additional capital expenditures of 3.1 million USD (0.06 USD $\cdot \ \mathrm{kg}^{-1}$ of BOG per year) and an energy demand of 6.33 $\mathrm{kWh}_{el} \cdot \mathrm{kg}_{BOG}^{-1}$. Other estimates in the literature assume capital costs (CCs) of 6.85--7.42 USD $\cdot \ \mathrm{kg}^{-1}$ of BOG per year and energy demand of 3.3--11.0 $\mathrm{kWh}_{el} \cdot \mathrm{kg}_{BOG}^{-1}$ \cite{Petitpas.2018, Hyunyong.2019} for re-liquefaction.

\section{Metrics for energy supply chain assessment: general formulation}
\label{sec:formulation}

In order to analyse and compare the ECs from production to utilisation, three assessments with corresponding metrics are utilised: 
\begin{enumerate}
    \item Energetic assessment: thermodynamic system efficiency,
    \item Environmental assessment: combined \ce{CO2} intensity,
    \item Economic assessment: LCOE (levelized cost of electricity).
\end{enumerate}

According to the schematic shown in Fig.~\ref{fig:TransportDiagram}, the combined formulation to evaluate these metrics considers the following steps of the supply chain: production, liquefaction (if applicable), short-term storage at the export terminal (intermediate storage), long-distance shipping, long-term storage at the import terminal, and the utilization in a power plant. Short-distance transport is out of the scope of the present work since it requires a detailed assessment of many locations and routes, both between producers and export ports, and import ports and power plants.

All calculations were carried out considering the material transported by a single cargo ship, with its cargo capacity fully occupied by the selected EC. Boil-off of LNG and L\ce{H2}, as well as other energy losses, affect the content of the cargo during the trip and the storage. Only \ce{CO2} emissions associated directly with each step of the process are considered, with indirect emissions associated with the electricity generation taken from estimates available in the literature.

\subsection{Energy balance and efficiency}
The relative share of energy losses for each step in the supply chain can be calculated by the ratio of the energy required or dissipated in each step and the total energy input of the system. The total energy input of the supply chain can be calculated according to:
\begin{equation}
   W_{total} = W_{prod} + W_{liq}+ W_{store_1} + W_{trans} + W_{store_2} \ [\mathrm{kWh}] \,,
   \label{eq:w_total}
\end{equation}
using equations \ref{eq:WprodfromPE}, \ref{eq:W_Liq}, \ref{eq:W_storefromMEC}, and \ref{eq:W_LD} of the Appendix. Therefore, the share of the total required energy in each step is given by:
\begin{equation}
   \epsilon_{i} = 100 \cdot \frac{ W_{i}}{W_{total}} \ [\mathrm{\%}] \,.
   \label{eq:e_loss}
\end{equation}

The thermodynamic system efficiency is given by the ratio of electrical energy output to the total energy input:
\begin{equation}
   \eta = 100 \cdot \frac{W_{elec}}{W_{total}}\ [\mathrm{\%}] \,.
   \label{eq:efficiency}
\end{equation}

\subsection{Carbon dioxide emissions}

The \ce{CO2} emissions can be considered as the sum of the contribution of each stage, as calculated using equations \ref{eq:CO2prodfromMEC}, \ref{eq:CO2_liq}, \ref{eq:CO2_store}, \ref{eq:CO2_LD}, and \ref{eq:CO2_elec} of the Appendix. For processes where methane (\ce{CH4}) is emitted, the equivalent in \ce{CO2} intensity, calculated by the relation between the global warming impact of methane and that of \ce{CO2}, is also accounted for. If given on the basis of electric energy output, the combined equivalent \ce{CO2} emitted can be calculated as:
\begin{equation}
   E_{\mathrm{\ce{CO2}}_{total}} = \frac{E_{\mathrm{\ce{CO2}}_{prod}} + E_{\mathrm{\ce{CO2}}_{liq}} + E_{\mathrm{\ce{CO2}}_{store_1}} + E_{\mathrm{\ce{CO2}}_{trans}} + E_{\mathrm{\ce{CO2}}_{store_2}} + E_{\mathrm{\ce{CO2}}_{elec}}} {W_{elec}} \ [\mathrm{kg}_{\mathrm{\ce{CO2}} \cdot \  \mathrm{kWh}^{-1}}] \,.
   \label{eq:CO2_global}
\end{equation}

\subsection{Levelized costs of electricity}
Relating costs to the electric energy output, the LCOE can be calculated as the sum of the cost contributions of each stage:
\begin{equation}
   C_{total} = \frac{C_{prod} +  C_{liq} + C_{store_1} + C_{trans} + C_{store_2} + C_{elec}}{W_{elec}} \ [\mathrm{USD} \cdot \mathrm{kWh}^{-1}] \,.
   \label{eq:C_global}
\end{equation}

In the analysis presented, a carbon tax $ T_{\mathrm{\ce{CO2}}}$ is taken into account by adding it to the total costs, multiplied by the amount of \ce{CO2} emitted in the relevant stage. It is assumed that the carbon tax only applies to direct emissions from the power plant (utilization step) and does not extend to \ce{CO2} emissions resulting from other steps along the process chain, such as transport, mining, or drilling for the conventional ECs.
\begin{equation}
   C_{total,CT} = C_{total} + \frac{{E_{\mathrm{\ce{CO2}}_{elec}} \cdot T_{\mathrm{\ce{CO2}}}}}{W_{elec}} \ [\mathrm{USD} \cdot \mathrm{kWh}^{-1}] \,.
   \label{eq:C_global2}
\end{equation}

Even if an EC's energy efficiency and emission profile is suitable, it must achieve a competitive cost of electricity for its successful adoption. Therefore, the equations for the costs are shown in more detail. For a reasonable estimation of the LCOE, the main aspects involved in every stage of the process must be considered, which include capital expenditures (CAPEX) and operating expenses (OPEX). 

\subsubsection{CAPEX and OPEX}
The annuity method is a well-established approach for evaluating projects from an economic perspective due to its ease of use and transparency. It calculates equal annual payments based on the present value of the initial investment costs. The calculation involves utilizing a capital recovery factor (CRF) and a constant discount rate $i$ over the project's economic lifetime $n$
\begin{equation}
   CRF = \frac{i(1 + i)^n} {(1 + i)^n-1} \,.
   \label{eq:CRF}
\end{equation}

The CAPEX corresponds to the yearly annuity based on the specific capital costs (CC in $\mathrm{USD} \cdot \mathrm{kg}_{EC}^{-1}$) of acquired equipment facilities, technology and other assets. In this sense, the capital costs are multiplied by the CRF to determine equal annual payments:
\begin{equation}
   CAPEX = CRF \cdot CC_{EC} \ [\mathrm{USD} \cdot (\mathrm{kg}_{EC} \cdot year)^{-1}] \,.
   \label{eq:CAPEX_year}
\end{equation}

The fixed OPEX includes the yearly rent, wages and maintenance costs, among others (note that fuel, electricity, and Suez Canal costs for ships are calculated separately). It is usually given as the fraction $\gamma$ of the (specific) CC:
\begin{equation}    \label{eq:OPEX}
   OPEX = \gamma\cdot CC_{EC}\ [\mathrm{USD} \cdot (\mathrm{kg}_{EC} \cdot year)^{-1}]  \,.
\end{equation}

\subsubsection{Cost of production}

Conventional production plants can operate almost all year round, whereas production plants utilizing renewable energy are subject to
variable renewable electricity full load hours due to volatility. Consequently, the installed capacity of plants utilizing renewables must be larger than the nominal capacity of downstream plants (such as the liquefier in the case of \ce{H2}). This fact is usually considered with a capacity factor (CF), defined as the mean plant operation times divided by the maximum operating hours. The corresponding CAPEX and fixed OPEX have to be adjusted with the CF. The cost of the process can be calculated considering the CAPEX, the OPEX incorporating the corresponding CF and the cost of the feedstock:
\begin{equation}    \label{eq:CprodfromMEC1}
   C_{prod} = m_{EC} \cdot (CC_{EC}\cdot\frac{CRF+ \gamma_{prod}}{CF_{prod}}+ C_{\mathrm{feedstock}}) \ [\mathrm{USD} \cdot  year^{-1}] \,.
\end{equation}

In this study, the production capacity factor is assumed to only apply to the electrolyzer and not to subsequent steps in the process chain for green ECs (i.e. hydrogen liquefier, shaft furnace). To allow the continuous operation of these subsequent process steps, intermediate hydrogen storage and renewable energy could be required (not explicitly considered here). Furthermore, the CAPEX, OPEX and conventional fuel feedstock are considered the same in all the production countries for each EC.

For the cases in which the feedstock is only electricity, it can be calculated as a function of the electricity cost:
\begin{equation}    \label{eq:CprodfromMEC2}
   C_{feedstock} = \frac{W_{prod}}{m_{EC}} \cdot C_{elec} \ [\mathrm{USD} \cdot \mathrm{kg}^{-1}] \,.
\end{equation}

\subsubsection{Cost of liquefaction}

The costs associated with liquefaction can be given by the CAPEX, OPEX and the associated electricity input and costs:
\begin{equation}    \label{eq:CAPEX_C_liq}
   C_{liq} = m_{EC} \cdot (CC_{liq}\cdot (CRF+ \gamma_{liq}) + C_{elec} \cdot w_{liq}) \ [\mathrm{USD} \cdot  year^{-1}] \,.
\end{equation}

\subsubsection{Cost of storage}

The storage costs involve the electricity input for maintaining the EC in transportable form (i.e. re-liquefaction of the BOG for NG and \ce{H2}.) Furthermore, the CAPEX and OPEX are only considered proportionally to the operation time $t_{store}$. Therefore, the total cost of the storage can be computed from:
\begin{equation}    \label{eq:C_store}
   C_{store} = m_{EC} \cdot (\frac{CC_{store} \cdot t_{store}}{t_{year}} \cdot (CRF+ \gamma_{store}) + C_{elec} \cdot w_{store} \ [\mathrm{USD} \cdot  year^{-1}] \,.
\end{equation}

\subsubsection{Cost of long-range transport}

Long-range transport costs are calculated based on the fuel consumed by the ship, minus the amount of which that comes from the boil-off of the EC when used as fuel during the one-way trip, along with the CAPEX and OPEX, on a USD per day basis, and the costs of passage through the Suez, if they are used during the trip. Analogously to the storage, the total operation days per year ($\mathrm{t}_{trans}$) are considered:

\begin{multline}    \label{eq:C_LD}
   C_{trans} = \frac{{t}_{trans}}{{t_{year}}} \cdot (CAPEX_{trans} + OPEX_{trans}) + \\  2 \cdot C_{canal} + C_{fuel} \cdot \frac{W_{trans} - (m_{cargo} - m_{EC_{trans}}) \cdot LHV_{EC}}{LHV_{fuel}} \ [\mathrm{USD} \cdot  year^{-1}] \,.
\end{multline}

For a more detailed calculation of the variables involved in ($\mathrm{C}_{trans}$), readers are referred to the Appendix.

\subsubsection{Cost of electricity generation}

The total costs of the power plant operation are given by the sum of the CAPEX and the OPEX, considering its operating hours through a capacity factor. No additional fuel costs are accounted for since the cost for the synthesis, transport and storage are allocated to the preceding processes:
\begin{equation}    \label{eq:C_elec}
   C_{elec} = W_{elec}  \cdot (CC_{elec}\cdot\frac{CRF+ \gamma_{elec}}{CF_{elec} \cdot 8760h}) \ [\mathrm{USD} \cdot  year^{-1}] \,.
\end{equation}

\subsection{Sensitivity analysis}

Evaluating the influence of a certain parameter upon the overall LCOE, a sensitivity analysis is performed according to:
\begin{equation}
   S_{f,x} = \frac{x_i}{f(x_i)} \cdot \frac{\partial f}{\partial x} \Bigr\rvert_{x_i} \,,
   \label{eq:sensitivity}
\end{equation}
where $S_{f,x}$ refers to the sensitivity of the target quantity $f$ with respect to the input quantity $x$. The sensitivity is determined in a discrete way, using a defined perturbation of an input variable to evaluate its influence on the result:
\begin{equation}
   S_{f,x} = \frac{x_i}{f(x_i)} \cdot \frac{f(x_i+\Delta x) - f(x_i)}{\Delta x} \,.
   \label{eq:sensitivity_d}
\end{equation}

If the variation is defined as a fraction of the base input variable, Eq.~(\ref{eq:sensitivity_d}) can be reformulated as:
\begin{equation}
   S_{f,x} = \frac{f(x_i\cdot (1 + R_{sens})) - f(x_i)}{R_{sens} \cdot f(x_i)} \,.
   \label{eq:sensitivity_d2}
\end{equation}

\subsection{Uncertainty analysis} \label{sec:uncertainty_analysis_theory}

As expected, since many estimates refer to novel technologies, the values of the variables used in the present work vary substantially depending on the data source. Along these lines, the calculations would benefit from uncertainty analysis. This study aggregates input quantities and parameters from literature research and online databases. Due to the sample's inherent bias, the input quantity's standard deviation, $\sigma$, is calculated using only the highest and lowest values encountered. Nevertheless, the approach shall provide trends and insights into the relative uncertainty $\sigma_{f,x_i}$ of the metric $f$, which originates from the uncertainty of the input quantity $x_i$. The relative uncertainty is computed as:

\begin{equation}
   {\sigma_{f,x_i}} = \frac{\partial f(x_i)}{\partial x_i} \cdot \sigma_{x_i}\,,
   \label{eq:reluncertainty}
\end{equation}
where the first term on the right-hand side represents the partial derivative of $f$ with respect to $x_i$ and $\sigma_{x_i}$ represents the standard deviation of the input quantity $x_i$. 

In order to approximate limits for the metric $f$, best- and worst-case scenarios are evaluated using the most favorable and adverse reported estimates of the input quantities $x_i$, respectively.

\section{Case study}
\label{sec:case_study}
This section presents a case study, considering the potential EC trade between selected clean energy exporters and importers. Routes from Morocco, Saudi Arabia and Australia to Germany are considered, as well as a route between Australia and Japan, using the formulation presented in this work.

\subsection{Importers of ECs: Germany and Japan}
\begin{figure}[!h]
		\begin{center}
        \includegraphics[width=0.9\textwidth]{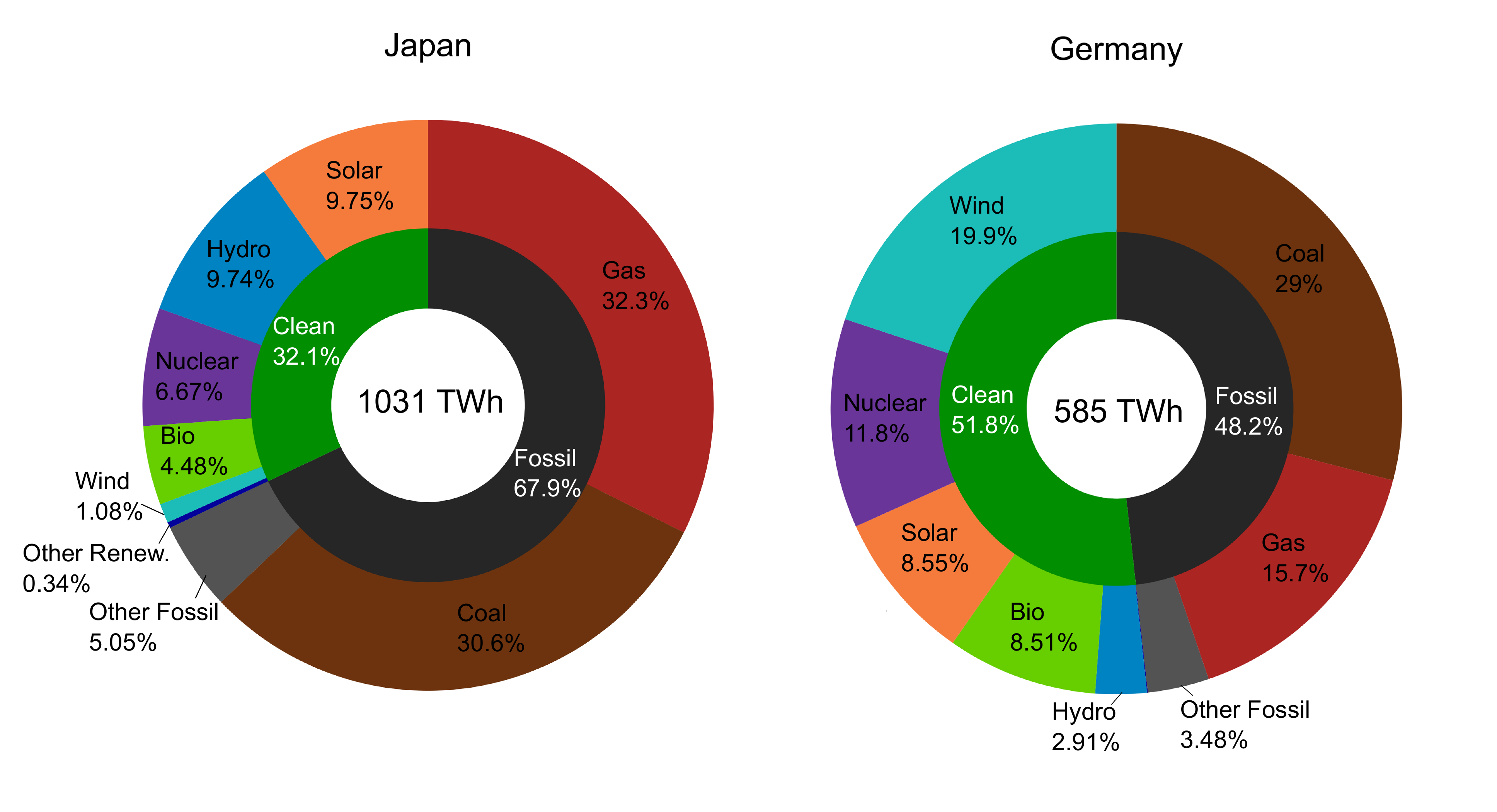}
        \end{center}

        \caption{Electricity mix in Japan and Germany \cite{emberNew}.}
		\label{fig:EnergyMixDEvsJP}
\end{figure}

Germany and Japan are two of the largest energy importers in the world~\cite{WEO2021}. Both countries are global economic powers ranking as the fourth and third biggest economies worldwide, respectively~\cite{worldbank}. At the same time, they lack sufficient natural resources to supply their energy demands. After the Fukushima incident, Germany committed to phase-out nuclear power plants~\cite{fukushima2011} and increased the electricity generation from fossil fuels to compensate for the reduced energy supply~\cite{emberNew}. Despite increasing solar and wind power generation in recent years, both Japan and Germany still strongly rely on fossil fuels, with shares of \SI{67.9}{\%} (Japan) and \SI{48.2}{\%} (Germany) of the total electricity generation~\cite{emberNew}, see Fig.~\ref{fig:EnergyMixDEvsJP}. Aiming to reduce \ce{CO2} emissions, both countries have stated policies to increase the share of renewable energy sources and consider importing green ECs in the long term~\cite{SRU2021,japan.2021,japan.2022}. Political stability and good commercial relations with countries in Asia, the Middle East, and Northern Africa allow the establishment of trade partnerships with net energy exporters.

Similarly to other European countries, Germany reduced its supply of NG due to the recent Ukrainian crisis and resorted to the reactivation of several coal power plants to stabilize its energy supply while taking into account a temporal relapse in terms of \ce{CO2} emissions. The country seeks clean and secure energy solutions for its economy in the long term. The recent adjustments in the energy supply underline that its geographical location allows various ECs to be imported via land and sea. On the contrary, Japan's energy imports must arrive primarily via sea routes, often over long distances, requiring stable and efficient ECs suitable for such trips.

\subsection{Exporters of ECs: Morocco, Saudi Arabia and Australia}
Morocco, Saudi Arabia, and Australia have been chosen as potential energy exporters due to their abundant sun and wind, which entails a potential for producing considerably more electricity than what is demanded nationally. These countries are reportedly pursuing projects on renewable electricity production and have increased their share of renewables in the energy mix in recent years~\cite{emberNew}. Exporting the surplus energy to distant countries requires the electricity to be converted and stored employing suitable ECs. In order to avoid \ce{CO2} emissions in the supply chain downstream, zero-carbon ECs are pursued

As of today, Morocco is still a net energy importer~\cite{IEAwebsite} and strongly relies on coal for its electricity supply (\SI{58.8}{\%} of its annual electricity mix~\cite{emberNew}). However, the country has pledged to significantly increase its renewable electricity capacity by 2030, with 2.2\,GW for wind power and 4.0\,GW of solar. Risks for such initiatives stem from the country's low investment capacity, which makes it dependent on international partners to achieve its goals~\cite{unfcc2021}. Morocco has intense trade relations with the EU, its leading commercial partner, and it is already linked to the European NG grid through the Maghreb-Europe Gas Pipeline~\cite{endfossil2022}. Due to its political stability, partnering the country's energy potential with the EU's investment capacity could lead to a rapid scale-up of renewable energy supply, storage, and export, mutually increasing energy security.

Saudi Arabia is one of the world's largest fossil fuel exporters. As of 2021, its electricity mix is based on NG by \SI{61}{\%}, on oil-derived fuels by \SI{38.6}{\%}, with solar power only accounting for \SI{0.5}{\%} and other sources less than \SI{0.1}{\%} \cite{emberNew}. However, due to its extensive arid territory and unique geographical position, it exhibits immense potential for solar-based electricity generation year-round. In 2020, photo-voltaic solar power (PV) reached a record low of USD 0.0104\,/kWh in the country which hints at the high potential for an economically viable production of ECs~\cite{IRENA2020}. Overall, the country's economy is considered as \textit{high-income}~\cite{eeas} with an established infrastructure for trading bulk goods. In the future, solar-fueled production of zero-carbon ECs will allow Saudi Arabia to diversify its portfolio as an international green energy exporter.

Australia is an important energy exporter, with net annual values of 2923\,TWh of coal and 1149\,TWh of NG exported~\cite{IEAwebsite}. Of the 244.6\,TWh of electricity produced in the country in 2021, \SI{51.3}{\%} was harnessed from coal, \SI{17.9}{\%} from NG and \SI{10.9}{\%} from solar power~\cite{emberNew}. Solar and wind electricity are the fastest-growing sectors in their energy mix. Australia also has an enormous potential to generate clean energy from these sources year-long due to the vast arid regions. Apart from that, the country has the world's largest iron ore reserve (51\,Gt). In 2020, it was the biggest producer of iron ore, with a yearly production of 0.92\,Gt, roughly \SI{37}{\%} of the total global supply~\cite{mcs2022}. The Pilbara region, in the West of the country, presently accommodates both the largest iron ore mining facilities in the country~\cite{bhpcorridor} and the location of the Asian Renewable Energy Hub~\cite{areh}. The latter is an ambitious project for PV and wind electricity generation with 26\,GW combined capacity, of which up to 23\,GW will be dedicated to producing green hydrogen and ammonia for export. This geographical location would also be ideal for a reduction site in a circular iron economy, enabling the production and recycling of green iron at scale, with the iron ore infrastructure of Port Hedland available for long-distance trade.

\subsection{Trade routes} \label{sec:routes}

\begin{figure}[h]
		\begin{center}
        \includegraphics[width=0.8\textwidth]{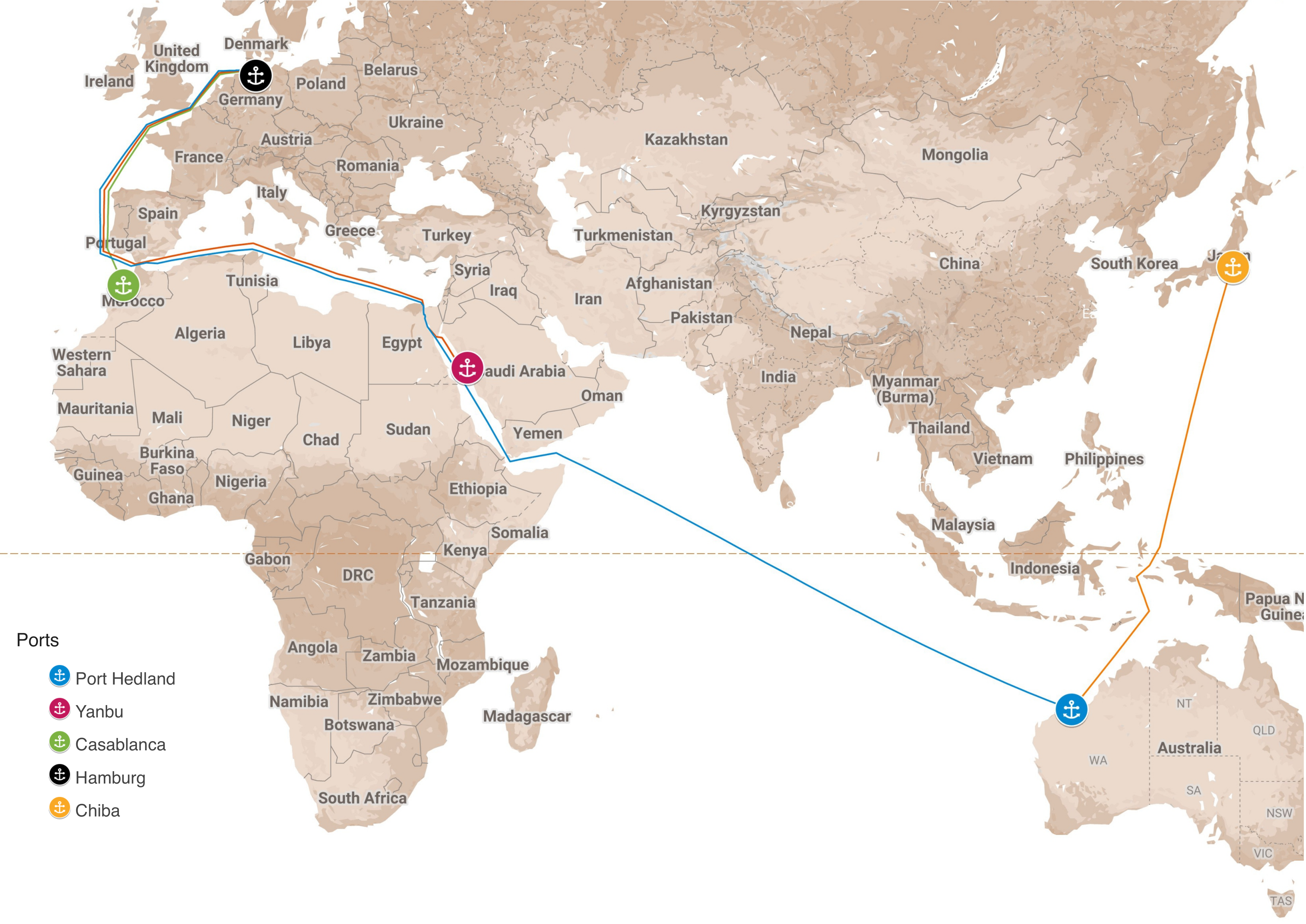}
        \end{center}
        \caption{Maritime trade routes from Morocco, Saudi Arabia and Australia to Germany and Japan.}
		\label{fig:RouteMap}
\end{figure}

Trade routes are chosen between important seaports of the energy exporters and importers, see Fig.~\ref{fig:RouteMap}. For the green energy importing countries, Germany and Japan, the clusters of seaports of Hamburg and Chiba are selected, respectively. These ports are assumed to provide sufficient capacity for processing and redistributing large amounts of ECs.

As a potential green energy exporter, Morocco can supply ECs to Europe without using the Suez Canal, which is a cost-advantage compared to other routes. The route to Germany by sea is the shortest among all Northern African countries, with a distance of 3098\,km between the ports of Casablanca and Hamburg. To the West, Saudi Arabia has access to the Red Sea, using the Suez Canal to transport goods to the Mediterranean region and Europe. Its route to Germany covers a distance of 7686\,km between the ports of Yanbu and Hamburg, which can be considered a medium-range distance in this comparison. 

The Pilbara region in the Northwest of Australia can be connected to Europe via the Suez Canal, with a distance of 18172\,km from Port Hedland to Hamburg. To Japan, the route between Port Hedland and Chiba results in a travel distance of 6704 km. All trade routes are evaluated for fossil and renewable energy carriers. Thus, even though not being typical coal exporters, the hypothetical shipping of coal from Morocco and Saudi Arabia is included for mathematical comparison.

\subsection{Variables and assumptions: costs, \ce{CO2} emissions, energy efficiencies} \label{sec:assumptions}

In order to calculate the costs, emissions and efficiency of each step of the supply chain, careful quantification of the variables must be carried out. Given the wide range of the variables incorporated in this model, the calculated results refer to the average between best- and worst-case scenarios based on the values that yield the lowest/highest costs, efficiency, and \ce{CO2} emissions, respectively. The best- and worst-case results are indicated using error bars. Here, values are collected from the literature and combined with assumptions per the state-of-the-art technologies. 

\textbf{Global variables:} Several global variables, which are used in the analysis, are defined in Table~\ref{table:global_var} (assumptions marked by footnotes). 
All equipment is considered to have an economic lifetime of 20 years. Assuming a interest rate of \SI{5}{\%}, the fixed CRF can be calculate with Eq.~\ref{eq:CRF} to \SI{8}{\%}. Fuel prices are set conservatively to the range indicated in Table~\ref{table:FF_production} for the Rotterdam trade hub and the Henry Hub for coal (8.0-43.7\,$\mathrm{USD} \cdot \mathrm{MWh}^{-1}$) and NG (8.8-18.5\,$\mathrm{USD} \cdot \mathrm{MWh}^{-1}$), respectively. Coal properties are taken from the averaged values presented in Table~\ref{tab:EC_properties}.

While market prices for electricity can vary broadly according to time and location, they are chosen according to present-day values and expectations referring to the range given by IRENA~\cite{Irena.2022b}.

\begin{table}[H]\scriptsize
\caption{\textbf{Global variables used in the case study.} }
\centering
\begin{tabular}{l c c c}
    \hline
     Global variable & Unit & Range & Ref.\\
    \hline
     € to USD rate & [$\mathrm{EUR} \cdot \mathrm{USD}^{-1}$] & 1.05$^1$  & \cite{EuropeanCentral.22} \\
     Clean electricity cost& [$\mathrm{USD} \cdot \mathrm{kWh}_{el}^{-1}$] & 0.01-0.1  &\cite{Fraunhofer.2021,IRENA2020,Irena.2022b} \\
     Conventional electricity cost & [$\mathrm{USD} \cdot \mathrm{kWh}_{el}^{-1}$] & 0.05-0.19  & \cite{Fraunhofer.2021,Irena.2022b} \\
     \ce{CO2}-emissions for conventional electricity & [$\mathrm{kg}_{\ce{CO2}}\cdot \mathrm{kWh}_{el}^{-1}$] & 0.475 &  \cite{IEAwebsite}  \\
     Carbon tax & $[\mathrm{USD}\cdot \mathrm{t}_{\ce{CO2}}^{-1}]$ & 85.7 $^2$ & \cite{Ember.2022}\\
     OPEX as share of CC & [\%]& 4$^3$ & \cite{thefutureofhydrogen}\\
     Capital Recovery Factor (CRF) & NA & 0.08 & \cite{thefutureofhydrogen,JOHNSTON2022}\\
     \hline
     \multicolumn{4}{l}{$^1$Assumption, values observed by January 2023;} \\
     \multicolumn{4}{l}{$^2$market  observed by December 2022;} \\
     \multicolumn{4}{l}{$^3$used value unless otherwise stated.}     
\end{tabular}
\label{table:global_var}
\end{table}

\textbf{\ce{CO2} taxation:} To account for future \ce{CO2} taxation in Germany and Japan, 86\,USD$/\mathrm{t}_{\ce{CO2}}$ (90\,EUR$/\mathrm{t}_{\ce{CO2}}$) are assumed as a baseline for the utilization of ECs, which is based on the market value of the EU carbon credits. Later, a parametric variation of the carbon tax is carried out in a break-even analysis to demonstrate the impact of carbon taxation on EC supply chains due to evolving policies. The \ce{CO2}-intensity of the global electricity mix estimated by the IEA~\cite{IEAwebsite} is assigned to conventional electricity.

\textbf{EC production:} Variables related to the production of ECs (costs, \ce{CO2} emissions and efficiencies) are given in Table~\ref{table:production}. All values, except for the total costs for green ECs, have been obtained from literature and pertain to existing, planned, or conceptualized production facilities. Notably, the assumption is made that green iron is produced from recycled iron oxides, eliminating the need for feedstock costs. However, the required iron ore costs, especially if constantly recycled, are insignificant (as discussed in Sec.~S~2). The calculations for the production of green iron efficiency are based on those presented by Vogl et al. in reference~\cite{Vogl.2018}. The total specific costs for green ECs are determined by the specified variables.

\begin{table}[h]\scriptsize
\caption{\textbf{Ranges of crucial production-related characteristics}.The CC given for \ce{H2} and \ce{Fe} correspond to the electrolyzer and the shaft furnace with corresponding EAF, respectively. More detailed information are available within Table~S~2.} 
\centering
\begin{tabular}{ l c  c  c  c  c  c }
    \hline
     & Unit  & NG & Coal  & Green \ce{H2} & Green \ce{Fe} & References   \\
    \hline
   Efficiency & [$\%$] & NA & NA & 50-74&  46-60 &\cite{thefutureofhydrogen,IRENA2020,Vogl.2018,Bhaskar.2020}\\
   CF electrolyzer & [$-$] & NA & NA &  0.19-0.74 & 0.19-0.74 & \cite{thefutureofhydrogen,Fraunhofer.2021,Fasihi.2020} \\
    Specific CC &  [$\mathrm{USD} \cdot \mathrm{kW}_{el}^{-1} / t_{\mathrm{EC}}^{-1}$] & NA &  NA  &  450-1400 & 225-430 & \cite{thefutureofhydrogen, IRENA2020, Facchini.2021,IEA.2010, Woertler.2013} \\
    Total cost & [$\mathrm{USD} \cdot \mathrm{kg}_{\mathrm{EC}}^{-1}$] & 0.13-0.26 & 0.07-0.37 &  0.74-11.96$^1$ & 0.11-0.84$^1$ &\\
   \ce{CO2} emissions & [$\mathrm{kg}_{\ce{CO2}} \cdot \mathrm{kg}_{\mathrm{EC}}^{-1}$ ] & 0.66$^2$ & 0.06$^2$ &  0 & 0 & \cite{IEAwebsite,CommonwealthAustralia2019,thefutureofhydrogen,ISTR2020}  \\
    \hline
    \multicolumn{7}{l}{ $^1$includes CAPEX, OPEX and costs for renewable energy} \\
    \multicolumn{7}{l}{ $^2$ related to the extraction (mining, drilling) of the ECs} \\
\end{tabular}
\label{table:production}
\end{table}

\textbf{Liquefaction of NG and hydrogen:} Reference quantities for the liquefaction are given in Table~\ref{table:liquefaction}. NG and \ce{H2}-existing liquefaction specifications are based on the present technology, while for \ce{H2}-concepts, the assessment is based on projections from the literature. The latter is used for the evaluation.

\begin{table}[H]\scriptsize
 \caption{\textbf{Ranges of crucial liquefaction-related characteristics.} \ce{H2}-existing corresponds to state-of-the-art/existing liquefiers, and \ce{H2}-concepts correspond to potential future liquefiers. The latter is used within the case study. A compilation of literature values can be found in Table~S~3.} 
 \centering
 \begin{tabulary}{\linewidth}{l  c  c  c c c}
    \hline
     \multirow{2}{*}{} & \multirow{2}{*}{Unit}  & \multirow{2}{*}{NG}   & \multicolumn{2}{c}{\ce{H2}} &  \multirow{2}{*}{References}  \\
     & & & Existing & Concepts &\\
     \hline
    Specific energy demand &  $[\mathrm{kWh}_{el} \cdot \mathrm{kg}_{EC}^{-1}]$ & 0.24-3.75& 10-20 &4-13 &  \cite{Raj.2016,Vos.2020, DOE2019,Majid.2022,Ghafri.22, Majid.2022}\\
     Specific CC & [USD $\cdot \  {\mathrm{kg}_{EC}^{-1}}$] & 0.19-2.21 &8.6-14.3 & 3.21-17.1 & \cite{Raj.2016,Vos.2020, NCE.2016, DOE2019} \\
     OPEX as share of CC& [$\%$] &2-4 &  \multicolumn{2}{c}{2.5-4} & \cite{Raj.2016,Vos.2020,thefutureofhydrogen} \\
     \hline
 \end{tabulary}
 \label{table:liquefaction}
 \end{table}

\textbf{Long-distance transport:} For the present analysis, ships using conventional fuels for fossil ECs are compared to zero-carbon alternative ships for clean fuels. Further, all ships are chosen according to the maximum cargo capacity with the constraint that the overall ship dimensions allow a passage of the Suez Canal. The vessels identified for all ECs and their specifications are listed in Table~\ref{table:transportation}. While the fossil ECs NG and coal are already traded internationally via sea routes, this is not the case for \ce{H2} and iron\footnote{Although, it should be noted that direct reduced iron (DRI) is already being shipped internationally as briquettes or fines in the volume of 21.1\,Mt (2020) with regulations in place by the International Maritime Organization Code of Safe Practice for Solid Bulk Cargoes~\cite{midrex_2020}.}. Therefore, assumptions about future cargo ships transporting these ECs must be made here.

\begin{table}[h]\scriptsize
\caption{\textbf{Specifications of conventional and alternative vessels for EC transport.} Iron and coal cargo are measured in metric tonnes (90\,\% of DWT capacity), while the other ECs are measured in $\mathrm{m}^3$ (manufacturer's CBM). The LHV of HFO is considered that of Diesel (42.7\,MJ/kg)~\cite{MAN-6S70ME-C}. Fuel costs for HFO refer to the global average on 2022-05-17 for very low sulphur fuel oil from Ship \& Bunker~\cite{shipnbunker}. LNG and L\ce{H2} vessels use the boil-off gases (natural or forced) as the main fuel. CAPEX for the considered vessels are calculated using the correlations of Mulligan~\cite{mulligan2008}.}
\centering
\begin{tabular}{l c c c c c}

    \hline
     & Unit & NG  & Coal &  Green \ce{H2} &  Green \ce{Fe}\\
    \hline
    Vessel type & N/A & LNG carrier  & bulk carrier &  L\ce{H2} carrier &  bulk carrier \\
    Net Capacity & [t or $\mathrm{m}^3$] & 216000 & 160000 &  160000  &  160000 \\
    \multirow{2}{*}{Boil-off rate} & \multirow{2}{*}{[\% $ \cdot$ day$^{-1}$]} & 0.08--0.3 & \multirow{2}{*}{N/A}  &{0.1--0.4}&  \multirow{2}{*}{N/A}\\
      &   &  \cite{Kim.2019,Pospisil.2019} &  &{\cite{Vos.2020}} &\\
    Engine Power & [kW] & 39240 & 19620 &  39240 &  19620 \\
    Speed & [km $\cdot \mathrm{h}^{-1}$] & 36 & 26 &  36 &  26  \\
    Fuel type & N/A & NG & HFO &  \ce{H2}  &  \ce{H2} \\
    Fuel cost & [USD $\cdot \ \ {\mathrm{kg}_{\mathrm{fuel}}^{-1}}$] & 1.0 & 1.0 &   0 &  * \\
    Ship capital cost & [$10^6$ USD/ ship] & 236.0 & 78.3 &  283.2 &  78.3 \\
    OPEX & [\% of CC]&4&4&4&4 \\
    Suez trip cost & [USD] & 300000 & 300000 &  300000 &  300000  \\
    \ce{CO2}-emissions \cite{co2umwelt2016} & [$\mathrm{kg}_{\ce{CO2}} \cdot {\mathrm{kWh}_{\mathrm{fuel}}^{-1}}$] & 0.201 & 0.293 &  0 &   0 \\
    \multirow{2}{*}{Reference ship} & \multirow{2}{*}{N/A} & Kharaitiyat  & Genco Titus  &   Kharaitiyat  & {Genco Titus}  \\
    &  & \cite{HHIlist, MAN-6S70ME-C}& \cite{GencoTitus,  S70MC-C8.2-TII}&\cite{HHIlist, MAN-6S70ME-C} &  {\cite{GencoTitus,  S70MC-C8.2-TII}}\\
    \hline
    \multicolumn{6}{l}{*Costs of \ce{H2} calculated per route, combining the specific costs of production and storage of the fuel.}\\
\end{tabular}
\label{table:transportation}
\end{table}

As stated in Sec.~\ref{sec:maritimeTransport}, sea ships that can transport liquefied hydrogen at scale are unavailable as of 2023. Nevertheless, it can be expected that such ships will be commercially available in the upcoming decade, as verified by published roadmaps of cargo ship manufacturers~\cite{Kawasaki.2022,provaris} and one already existing pilot ship~\cite{Kawasaki.2019}. Thus, a future L\ce{H2} carrier is assumed to have 160000\,$\mathrm{m}^3$ of cargo, in line with the assumptions of Johnston et al.~\cite{JOHNSTON2022}, the IEA~\cite{thefutureofhydrogen}, and the next-generation Kawasaki L\ce{H2} ship~\cite{Kawasaki.2022}. Considering the necessary structural adaptations to maintain the temperature below -253\,\textdegree C during the trip, thermal insulation would have to be considerably more efficient than that of LNG ships, leading to a reduced cargo capacity and increased costs when compared to a similarly sized ship. Therefore, the ship is assumed to have the external dimensions of the reference LNG ship, 1.35 times lower volumetric capacity and $20\,\%$ more expensive. In Sec.~S~1 of the supplementary material, the effect of these assumptions on a potential \ce{H2} supply chain is studied by varying the hydrogen ship cargo capacity and costs.

Similar as for coal, a solid bulk carrier with comparable DWT capacity is chosen for iron transport. Even though iron transport demands certain safety measures, i.e.~sealed big bags or cargo containers, which require additional storage space, the limiting factor for storage will be the mass of the material\footnote{Note the density difference of a 4-5 factor between coal and iron.} While other zero-carbon fuels are conceivable for maritime transport, in the green scenarios for \ce{H2} and \ce{Fe}, it is assumed that the ships are fueled by \ce{H2}. Moreover, it is unlikely that a company operating a retrofitted power plant would also own the ships transporting iron and iron oxides. Considering the globalized transport of fossil fuels today, it can be assumed that a trading network of specialized companies would be established along the energy supply chain, likely increasing overall costs. The cost assessment utilizing the ship's CC and the CRF to annualize shipping costs should therefore be considered as an initial assessment to provide guidance. Similar approaches have been reported in the literature~\cite{JOHNSTON2022, Hank.2020}.

The ship's speed eventually leads to different transport times for the previously defined routes, as shown in Table~\ref{table:transport time}.

\begin{table}[h]\scriptsize
\caption{\textbf{Transport times} Round-trip times with respect to the defined trade routes and the transport speeds given in Table \ref{table:transportation} for the different ECs in days.}
\centering
\begin{tabular}{l  c  c  c c }
    \hline
   & LNG   & Coal  &  L\ce{H2} &  \ce{Fe} \\
    \hline
    CAS-HAM & 7.2 & 9.9&  7.2& 9.9\\
    YAN-HAM & 17.8 & 24.6&  17.8& 24.6\\
    HED-HAM & 42.1 & 58.2&  42.1& 58.2\\
    HED-CHI & 15.5 & 21.5&  15.5& 21.5\\
    \hline
\end{tabular}
\label{table:transport time}
\end{table}

\textbf{Storage facilities:} Storage specifications are shown in Table~\ref{table:store}. Intermediate storage is assumed to last for the round-trip duration (see Table~\ref{table:transport time}) with six additional days for loading and unloading the cargo. Long-term storage is fixed to 90\,days in order to account for seasonally varying demands for the ECs, which has to be paired with the trends in the availability of renewable energy supply (i.e. EC demand can be expected to increase during winter times). Table~\ref{table:store} shows the ranges of critical storage-related values, while a compilation of literature values can be found in Table~S~4.

\begin{table}[H]\scriptsize
\caption{\textbf{Ranges of crucial storage-related characteristics}. A compilation of literature values can be found in Table~S~4.}
\centering
\begin{tabular}{l c c  c c c c c}
\hline
\multirow{1}{*}{}  & \multirow{1}{*}{Unit}    & \multirow{1}{*}{LNG}   & \multirow{1}{*}{Coal}  &  \multicolumn{1}{c}{L\ce{H2}} &   \multirow{1}{*}{Iron} &  \multirow{1}{*}{References}  \\
    \hline
    \multirow{1}{*}{Boil-off rate} &\multirow{1}{*}{[wt.\%$\cdot \ \mathrm{day}^{-1}$]} & 0.0-0.15   & NA  & \multicolumn{1}{c}{0.04-0.40}  & NA & \cite{JOHNSTON2022, Pospisil.2019, Fesmire.2021, Vos.2020, EnergyGov.2017}\\
    \multirow{1}{*}{Specific CC}  &\multirow{1}{*}{[USD $\cdot \ \mathrm{kg}^{-1}$]} & 0.65-40  & $3.45 \cdot 10^{-4}$   & \multicolumn{1}{c}{11.95-148.80}  &  $ 3.45 \cdot 10^{-4}$ & \cite{JOHNSTON2022,NCE.2016, Rani.2010, Hank.2020} \\ 
     \multirow{1}{*}{OPEX as share of CC} & \multirow{1}{*}{[\%]} & 4  & 4   & \multicolumn{1}{c}{2-4}  &  4 &\cite{JOHNSTON2022, Vos.2020}\\
    \hline
\end{tabular}
\label{table:store}
\end{table}

\textbf{Power generation:} Using hydrogen and iron for power generation, the efficiencies are estimated to be close to those of NG (gas turbine) and coal power plants, respectively. For NG, the efficiency is estimated as that of a combined cycle gas turbine (CCGT) facility \cite{Fraunhofer.2021}, while being assumed as lower, at 50\%, for \ce{H2}\footnote{Extensive research is being carried out for \ce{H2}-fueled gas turbines. However, substantial technological development is still required to reach similar efficiencies as NG-CCGT.}. The capacity factor for the power plants is set to 0.06-0.34, in line with assumptions for future intermittence backup (medium to peak load power plants) \cite{Fraunhofer.2021}. The variables related to power generation are summarized in Table~\ref{table:utilization}.

\begin{table}[H]\scriptsize
\caption{\textbf{Overview of assumptions and references with respect to power generation from different ECs.} Values are based on \cite{Fraunhofer.2021, IEA.2021b,co2umwelt2016,Janicka2022}. More detailed information can be found in Table~S~10.}
\centering
\begin{tabular}{l  c c  c  c  c  c}
\hline
& Unit   & NG   & Coal  &    \ce{H2} &   {Iron}  \\
    \hline
    \multirow{1}{*}{Specific CC} & [USD$\cdot \mathrm{kW}_{el}^{-1}$] & 842-1158    & 1579-2316  &  CC\textsubscript{NG}  & (0.1-0.5 CC)\textsubscript{Coal} \\
     \multirow{1}{*}{Capacity Factor} & [-] & \multicolumn{4}{c}{0.06-0.34} \\
    \multirow{1}{*}{OPEX} & [USD$\cdot \mathrm{kWh}_{el}^{-1}$] & 21.05 & 23.16   & OPEX\textsubscript{NG}  &  OPEX\textsubscript{Coal}\\
      \multirow{1}{*}{Net efficiency} & [$\%$] & 60  & 46-50   & 50  &  46-50\\
      Direct \ce{CO2}-emissions  & [$\mathrm{kg}_{CO2} \cdot \mathrm{kWh}_{EC}^{-1}$]& 0.20 & 0.35 & 0 & 0 \\
      \hline
      \end{tabular}
\label{table:utilization}
\end{table}

\subsection{Results and discussion}\label{sec:results}

\subsubsection{Metrics for maritime transport of different ECs}

For quantifying the impact of various ship characteristics (e.g. size, fuel requirements, speed) on the supply chains of different ECs, reference data has been collected for different vessels. In particular, the relative energy demand and the specific capital costs are evaluated and analyzed. In order to facilitate comparison, both quantities are related to the transported energy that is chemically bound by the EC (measured in kWh$_\mathrm{EC}$).

For simplicity, the nominal speed of the ships is used, and unless otherwise specified, an engine efficiency of \SI{50}{\%} is assumed. In the following charts, vessel sizes are normalized with the maximum vessel size (in terms of net cargo) whose external dimensions still allow passage of the Suez Canal. The basis for normalization corresponds to the ships given in Table~\ref{table:transportation}. 

\begin{figure}[!h]
		\begin{center}
        \includegraphics[width=0.8\textwidth]{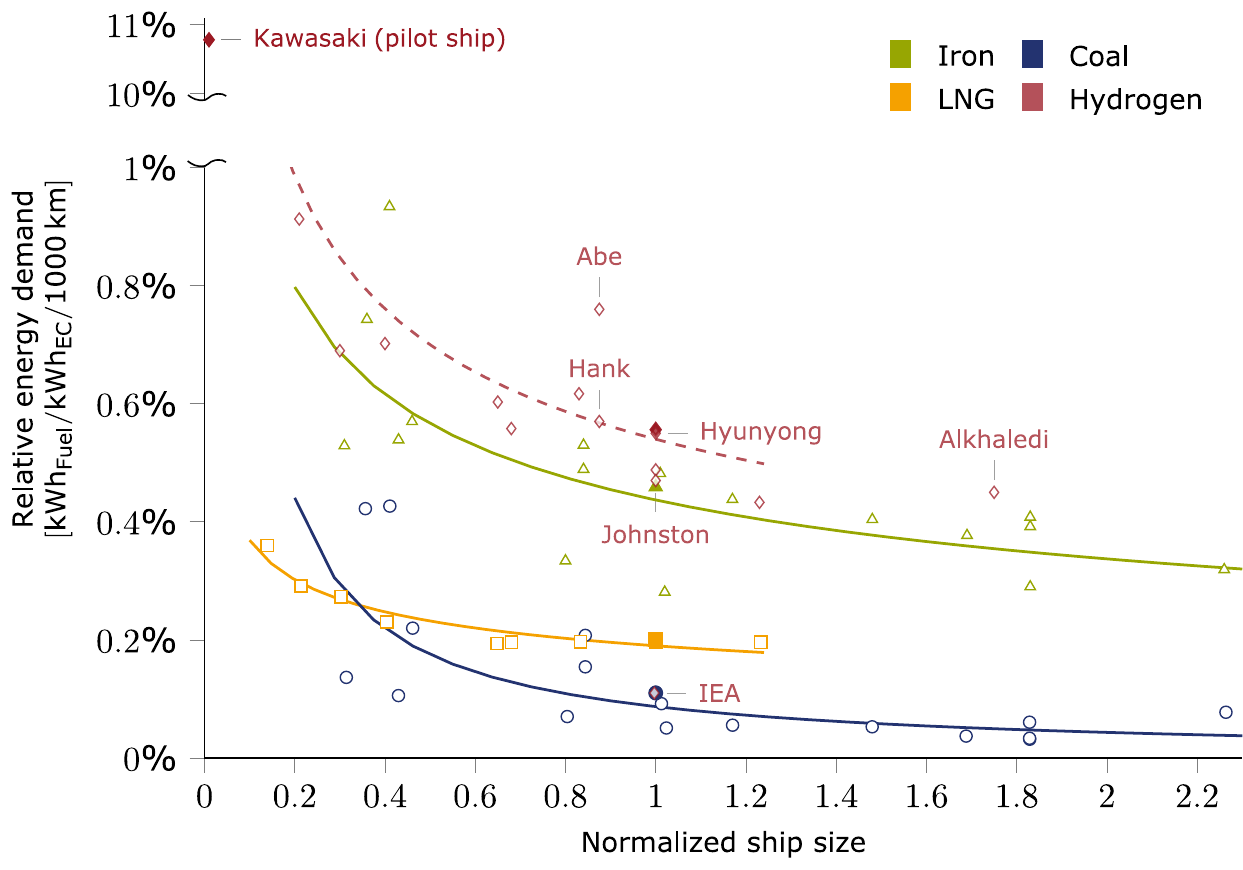}
        \end{center}
        \caption{Relative energy demand for selected vessels regarding their normalized size based on the Suez Canal limit (e.g. 1 corresponds to the maximal sizes suited for the Suez Canal). The lists of vessels are given in Table~S~5, S~6 and S~7. It is assumed that \SI{90}{\%} of the DWT of the solid bulk carriers is available for cargo. The net cargo volume for hydrogen ship is assumed to be \SI{0.75}{\%} of current LNG ships. Highlighted points correspond to (proposed) designs from the literature (Kawasaki~\cite{bairdmaritime}, Hank~\cite{Hank.2020}, IEA~\cite{thefutureofhydrogen}, Johnston~\cite{JOHNSTON2022}, Abe~\cite{Abe.1998}, Alkhaledi~\cite{Alkhaledi.2021}).}
		\label{fig:Relative vessel energy demand}
\end{figure}

Fig.~\ref{fig:Relative vessel energy demand} shows the relative energy demand which stems from moving a ship (kWh$_\mathrm{Fuel}$) in relation to the energy content of the cargo, per 1000\,km travelled. 
While the data for LNG and coal vessels correspond to real operating ships, it is assumed that the coal carriers can be repurposed to transport iron. As discussed in Sec.~\ref{sec:assumptions}, for L\ce{H2}, it is assumed that a hypothetical hydrogen carrier shows similarities to existing LNG vessels but with reduced net cargo capacity due to higher insulation requirements. As expected, the figure illustrates that the relative energy demand for all ship types decreases significantly with increasing ship size. While the vessels for the conventional ECs, coal and LNG, show relative energy demands of \SI{0.1}{\%} to \SI{0.2}{\%} per \SI{1000}{\km}, respectively, the transport of iron requires more energy ( \SI{0.4}{\%}-\SI{0.5}{\%} per \SI{1000}{\km} for medium and large vessels), which is directly linked to its lower gravimetric energy density compared to coal (see Table~\ref{tab:EC_properties}). In contrast to these operating (or repurposed) vessels, the largest operating L\ce{H2} vessel, which is a pilot ship, shows a relative energy demand of more than \SI{10}{\%}, which is prohibitive for long-distance transport. However, even if the technology could be scaled to reasonable \ce{H2} carriers sizes in the future, the relative energy requirement remains higher (\SI{0.5}{\%} to \SI{0.6}{\%} per \SI{1000}{\km} for medium and large ships) than for iron. As highlighted in Fig.~\ref{fig:Relative vessel energy demand}, this agrees with reference values for proposed designs documented in the literature. The assessment is only contrasted by the numbers reported by the IEA~\cite{thefutureofhydrogen} that assumes a more favorable transport efficiency for \ce{H2}, which is comparable to coal transport.

\begin{figure}[!h]
		\begin{center}
        \includegraphics[width=0.8\textwidth]{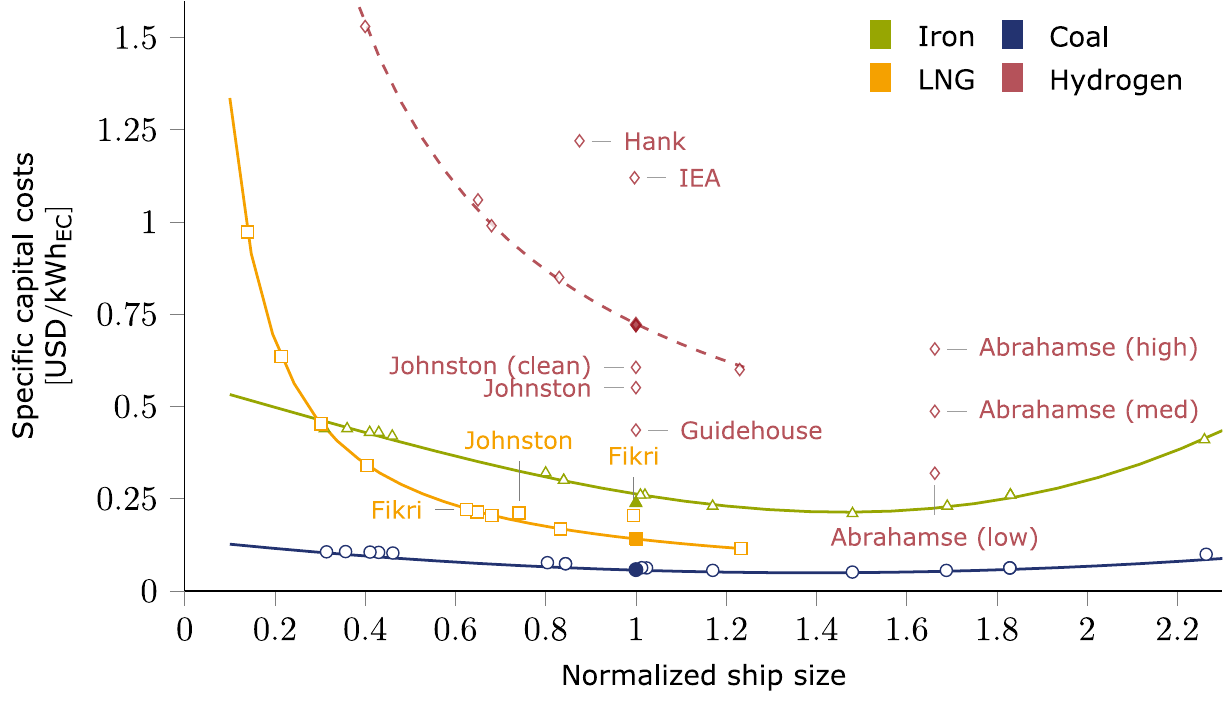}
        \end{center}
        \caption{Specific capital costs for selected vessel regarding their normalized size based on the Suez Canal limit (e.g. 1 corresponds to the maximal sizes suited to the Suez Canal). The lists of vessels are given in the supplementary material, Tables~S~5, S~6 and S~7 for which the capital costs are based on the correlations developed by Mulligan \cite{mulligan2008}. The diamonds correspond to literature values (Hank~\cite{Hank.2020}, IEA~\cite{thefutureofhydrogen}, Johnston~\cite{JOHNSTON2022}, Guidehouse~\cite{Guidehouse.2021}, Abrahamse~\cite{Abrahamse.2021}, Hyunyong~\cite{Hyunyong.2019}, Fikri~\cite{Fikri.2018}).}
		\label{fig:specific_vessel_CC}
\end{figure}

Fig.~\ref{fig:specific_vessel_CC} shows the specific capital costs for different ship types in terms of the energy content of the cargo. The results are based on the empirical correlations for LNG vessels and solid bulk carriers developed by Mulligan~\cite{mulligan2008} depicted in Sec.~\ref{sec:CC_ships} of the Appendix. While the prices for iron carriers are based directly on the estimates for solid bulk carriers, it is assumed that vessels capable of transporting L\ce{H2} are 20$\,$\% more expensive compared to LNG vessels, as previously discussed. Medium to large vessels for the conventional ECs coal and LNG show specific CCs of 0.15--0.3\,USD $\cdot \mathrm{kWh}_{EC}^{-1}$ and 0.05-0.15\,USD $\cdot \mathrm{kWh}_{EC}^{-1}$, respectively. Interestingly, the specific cost for very large bulk solid carriers increases again beyond a certain size, indicating limits for the economics of scale. Analogously to the relative energy demand of iron ships, the specific CCs of iron-carrying vessels are notably higher (0.25--0.5\,USD $\cdot \mathrm{kWh}_{EC}^{-1}$) compared to coal transporting vessels due to the gravimetric energy density differences. The cost assessments for future hydrogen vessels, as well as cost estimates reported in the literature (diamond symbols in Fig.~\ref{fig:specific_vessel_CC}), indicate even significantly higher costs for \ce{H2}-transport. It is emphasized that the estimates for \ce{H2}-vessels are subject to significant uncertainty, which is also indicated by the deviations among the literature values of over 100$\,$\%.

\FloatBarrier 

Long-distance maritime transport of conventional ECs is both energy- and cost-efficient, as evidenced by its decades of successful application worldwide. Even though the cost and relative energy requirements for transporting iron are higher compared to transporting conventional ECs, the present analysis shows that long-distance transport of iron via medium and large cargo ships is possible and will not exceed single-digit energy expenditures (in terms of the percentage of chemically-bound energy) even for very long distances such as Australia-Germany.

\subsubsection{Holistic energy supply chain analysis}

\begin{figure}[ht]
		\begin{center}
        \includegraphics[width=0.95\textwidth]{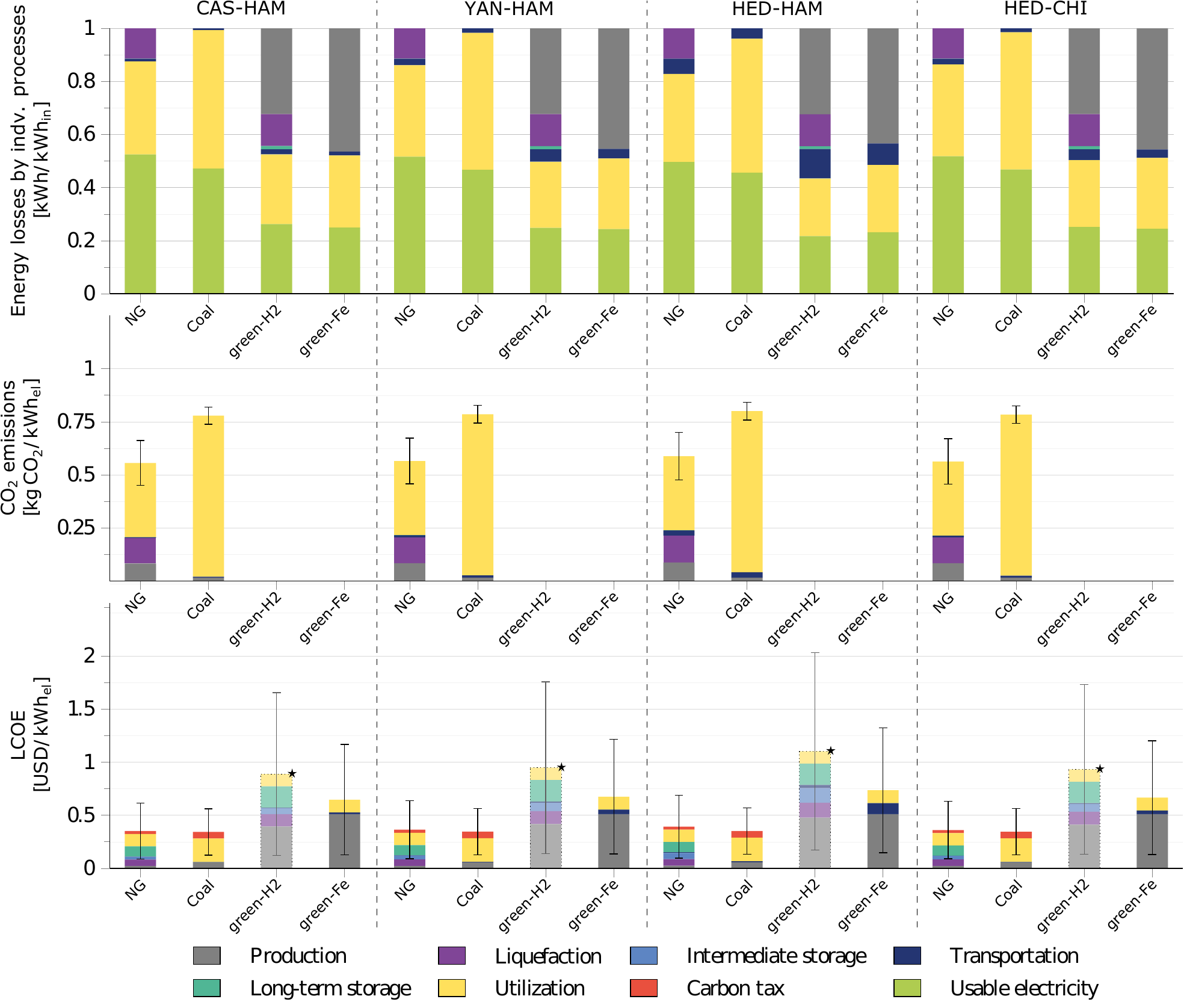}
        \end{center}
        \caption{Overview of the performance of the different ECs in terms of energy share by process (top), \ce{CO2} emissions (middle) and LCOE (bottom) for the specified routes. The corresponding process steps shown in Fig.~\ref{fig:TransportDiagram} are color-coded.}
		\label{fig:main_results}
\end{figure}

The combined results for the energetic assessment (thermodynamic system efficiency), the environmental assessment (\ce{CO2}-emissions), and the economic assessment (LCOE) of the EC supply chains are shown in Fig.~\ref{fig:main_results}. It is worth to notice that the cost assessment for hydrogen supply chains is shaded, indicating the significant uncertainty associated with its maritime transport, storage, and liquefaction. The cost estimate should therefore be understood as a hypothetical value. A more extensive exploratory cost assessment, which includes a variation of important transport parameters for the \ce{H2} supply chain, is provided in Sec.~S~1 of the supplementary material for interested readers.

\begin{figure}[!h]
		\begin{center}
        \includegraphics[width=0.95\textwidth]{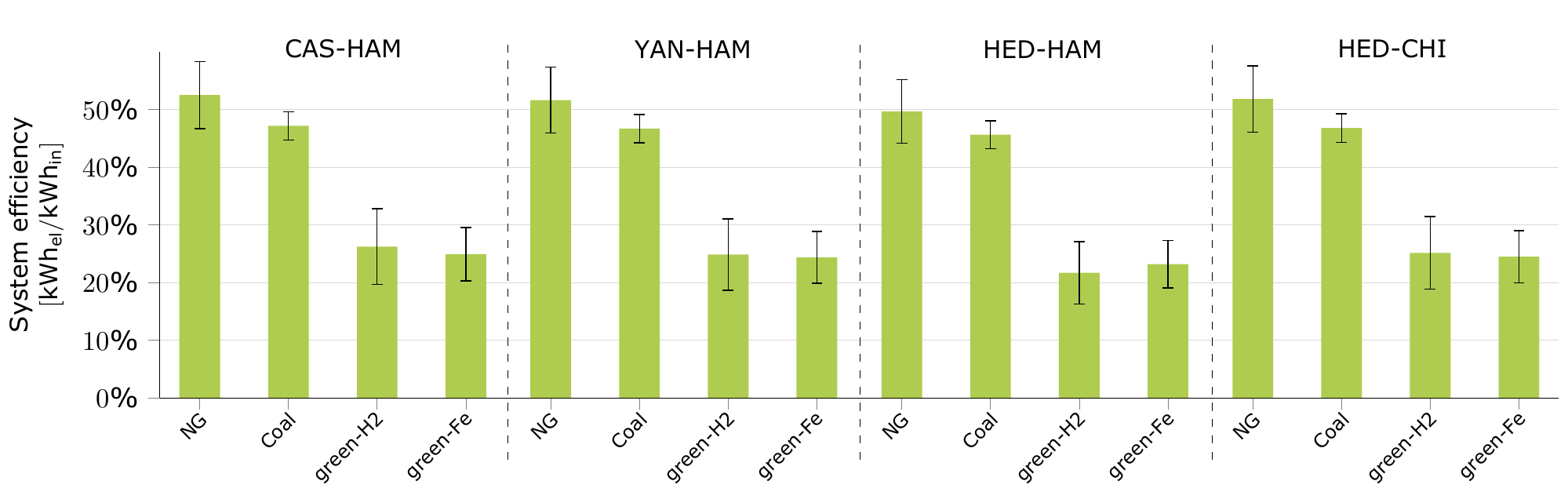}
        \end{center}
        \caption{Overall thermodynamic system efficiency of the energy supply chain for the selected ECs and routes.}
		\label{fig:system_efficiencies}
\end{figure}

Fig.~\ref{fig:main_results} (top) shows the energy losses for the individual processes along the energy supply chains of the ECs studied here. Notably, the energy share by process varies broadly between the ECs. Overall, coal and NG are the most energy-efficient ECs. However, given the differences in the process chains for conventional (primary energy without required production) and renewable ECs (secondary energy with required production), it is crucial to approach this comparison between conventional and renewable ECs with caution. Most energy losses for NG originate from its utilization process (i.e.~electricity generation) and liquefaction, with only minor losses due to storage and transport. As expected, the latter takes a higher toll on the efficiency when the ECs are transported over longer distances (due to maritime fuel consumption and longer intermediate storage times). On the other hand, coal does not need to be transformed to be transported and stored; it is also safe to be piled up in warehouses or yards. Therefore, the handling and processing of this EC require minor relative energy input. The main energy losses in the supply chain stem from the utilization, which is less efficient than that of NG. The impact of transport distance is also low for coal due to the high energy content that can be shipped per trip. 

The zero-carbon ECs, on the other hand, are very energy-intensive to produce, requiring an energy input that ranges from around \SI{30}{\%} to \SI{50}{\%} of the total energy expenditure of the considered supply chain. Green hydrogen further exhibits high energy losses due to liquefaction, transport, and storage since achieving and maintaining its liquid state at cryogenic temperatures significantly increases the energy demand for handling the EC. Even in its liquid form, L\ce{H2} transport is less efficient due to the small volumetric energy density of L\ce{H2}, but also due to the expected limitation in cargo capacity, which stems from the application of considerable thermal insulation. Consequently, the impact of the transport distance on the overall energy efficiency is comparatively high. Further losses are related to the boil-off (despite BOG being assumed to be used as fuel), which also has to be addressed for intermediate storage times. Iron is the most energy-intensive fuel to be produced. For the production and transport of conventional iron, fossil fuels are utilized, accounting for up to \SI{50}{\%} of the energy supply chain. Despite being less energy-intensive, green iron production relies on hydrogen for the reduction reactions, thus inheriting the production efficiency of the gas. Contrary to \ce{H2}, green iron storage and transport is very efficient. Due to the very high volumetric energy density of iron, high amounts of chemical energy can be shipped per trip, while its chemical stability facilitates efficient storage similar to coal. Nevertheless, material handling (packing, loading, unloading and storage) requires additional effort and sealable containers, such as super-bags, to avoid oxidation and ensure security. Overall, the impact of the distance on the energy expenditures for transport is lower than for \ce{H2} but higher than for fossil fuels due to the energy content delivered for each trip. 

The overall supply chain energy efficiency is shown separately in Fig.~\ref{fig:system_efficiencies} together with error bars corresponding to best-case and worst-case scenarios. As stated before, after extraction, transport, storage and utilization, NG and coal, are the most efficient ECs with \SI{46}{\%} - \SI{52}{\%} usable energy. For the zero-carbon ECs, \ce{H2} and iron, the holistic energy supply chain analysis yields efficiencies of \SI{23}{\%} - \SI{26}{\%} with either one being more efficient depending on the transport distance.

Assuming that all processes are optimized along the supply chains (upper bound of error bars in Fig.~\ref{fig:system_efficiencies}), the maximum potential efficiencies are \SI{32}{\%} and \SI{29}{\%} for green \ce{H2} and iron (assuming the route CAS-HAM), respectively. Note that short-range transport has not been considered here, which can influence the efficiencies of all ECs.

Comparing the \ce{CO2} emissions in Fig.~\ref{fig:main_results} (middle row), the largest emissions per kWh$_{\mathrm{el}}$ are produced from coal in the utilisation step, while other \ce{CO2} emissions along its supply chain are negligible small in comparison\footnote{It is worth noticing, however, that the coal source considered in this comparison is the open pit mining, which generates far fewer methane emissions than deep mining~\cite{ScienceI2021} which could change this assessment.}. In contrast, the utilisation of NG leads to lower \ce{CO2} emissions, but other emitting processes are the liquefaction\footnote{It is considered here that the electricity required for liquefaction is at least partially based on fossil fuels.} and the extraction process (\ce{CH4} emissions). In this case study, green hydrogen and iron, by definition, do not produce any \ce{CO2}, using fully decarbonised processes to produce, prepare, transport, store, and utilise the EC. In real scenarios, it is supposedly very challenging to avoid indirect emissions along the whole energy supply chain due to the production of machinery, infrastructure construction, and other auxiliary processes. It has to be stated, though, that the \ce{CO2} emissions from transport, independent of the route, have a minimal relative impact on the overall emissions compared to the very high \ce{CO2} intensity of conventional fossil fuels.

Analyzing the levelized costs for electricity (LCOE) in Fig.~\ref{fig:main_results} (bottom), it is found that contributions from different steps along the energy supply chain vary widely for different ECs. NG has the lowest share of fuel costs (production), which can be associated with the comparably low fuel prices for NG at the Henry Hub (see Table~\ref{table:FF_production}). However, capital costs and electricity expenses for the liquefaction and storage processes heavily affect the final price. NG utilization is also low-cost due to comparably low infrastructure and operation expenses. Coal has higher fuel costs but the least expensive storage, primarily due to easy handling and storage. When comparing the pre-utilization costs of coal and NG, coal shows overall lower costs. However, the use of coal results in high expenses for the power plant construction and operation, which make up the bulk of its supply chain costs. Especially due to the low capacity factors for the power plant ($CF=0.06-0.34$, medium to peak power plant), the high investment costs for the coal-fired power plant are dominant. Transport of both fossil fuels is also inexpensive due to the efficiency of their transport. On the other hand, \ce{H2} appears as the most expensive of all ECs, which can be attributed to the particularities of its supply chain, especially the high energy demand of the production process, the liquefaction and storage costs, which require advanced infrastructure and elevated electricity expenses. Due to the characteristics of \ce{H2}, the effect of transport distance on transport and storage is also significant. Firstly, the ship fuel consumption is affected, but, more importantly, the transport distance affects the follow-up costs from increased intermediate storage times and cargo losses from the usage on board (boil-off). Among the ECs evaluated, iron has the highest production costs. However, its advantageous transport and storage characteristics in comparison to hydrogen and the possibility of retrofitting existing coal-fired power plants give it a potential competitive advantage over hydrogen.
As expected, \ce{CO2} taxation (red bar for NG and coal in Fig.~\ref{fig:main_results}, bottom) negatively affects fossil fuel overall costs. However, the assumed value of 86\,USD/t$_{\ce{CO2}}$ in the utilization step only is not high enough to make green ECs economically competitive unless improvements on the green chains reach close to the best-case scenarios. The influence of the factors above is further investigated through a sensitivity analysis in the next section.

\subsubsection{Sensitivity analysis} \label{sec:sensitivity_analysis}

In this section, the relative impact of changes in the supply chain on the LCOE is investigated for the fossil fuels, NG and coal, and the potential zero-carbon alternatives, \ce{H2} and iron for the trade route between Germany and Australia (Hed-Ham). A positive perturbation of \SI{1}{\%} is imposed on selected variables, and the sensitivity of the LCOE is evaluated according to Eq.~(\ref{eq:sensitivity}). Thus, a sensitivity factor of -0.5 indicates that a \SI{1}{\%} increase in the respective variable leads to a \SI{0.5}{\%} reduction in the LCOE. The analysis reveals where improvements in the supply chain would more significantly affect the costs, representing important information for decision-makers and indicating where further development might be most beneficial.

\begin{figure}[h]
\begin{center}
\includegraphics[width=0.45\textwidth]{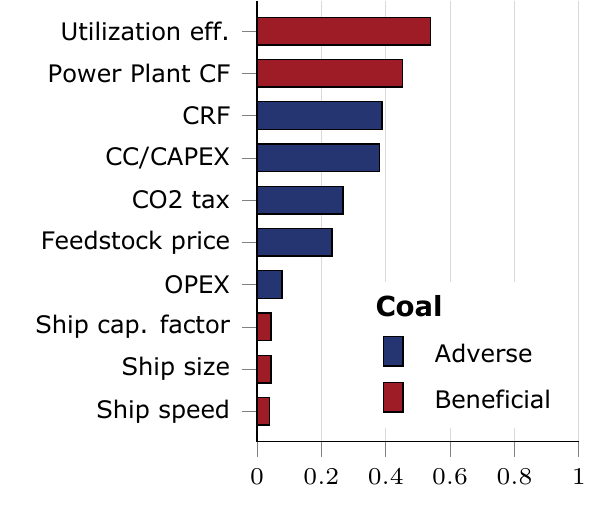}\quad%
\includegraphics[width=0.45\textwidth]{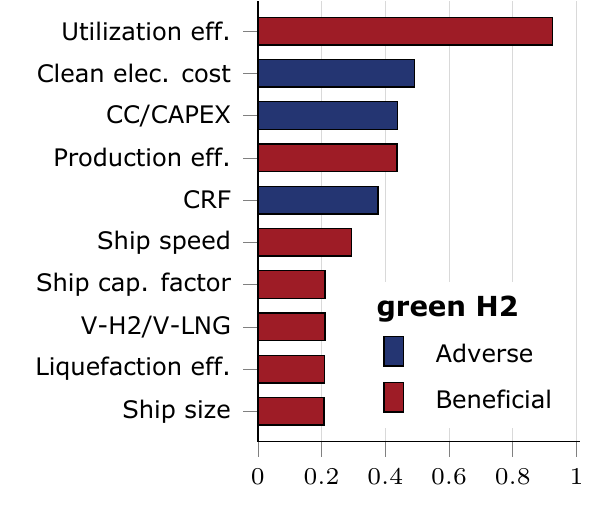}%
\end{center}
\begin{center}
\includegraphics[width=0.45\textwidth]{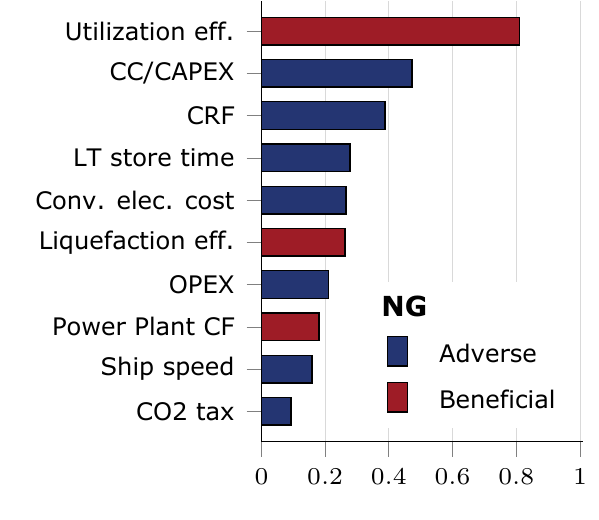}\quad%
\includegraphics[width=0.45\textwidth]{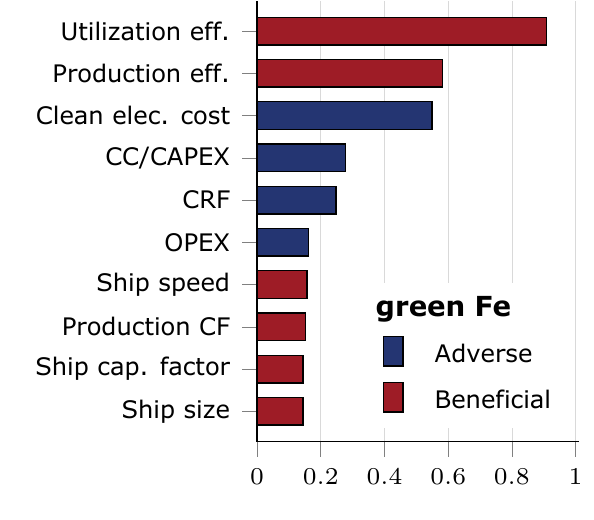}
\end{center}
\vspace{-6pt}
\caption{Sensitivity analysis with respect to the LCOE of the selected ECs. A negative sensitivity factor (e.g. \SI{-0.5}{\%}) indicates that a \SI{1}{\%} increase in the variable results in a beneficial decrease in the LCOE (e.g. \SI{0.5}{\%}).}
\vspace{-3pt}
\label{fig:sensitivity}
\end{figure}

Fig.~\ref{fig:sensitivity} shows the ten highest sensitivities calculated for the considered energy supply chains. For all ECs, the most influential variable for the LCOE is the utilization efficiency, which, when improved, leads to the most significant cost reduction. Furthermore, the capital costs of the infrastructure and the CRF show high sensitivity, which underlines the high investments associated with the required infrastructure. Particularly for coal, the power plant capacity factor (operation hours per year) significantly affects the LCOE due to the high capital costs of coal-fired power plants. Notably, the feedstock costs for coal are highly sensitive, while those for NG are not among the top ten sensitivities. This is due to the low reference feedstock cost of NG compared to coal. The LCOE of NG, on the other hand, is negatively influenced (hence, lowering the LCOE) by an increase in the liquefaction efficiency. Considering the liquefaction and storage demands for NG, an increase in storage time, which leads to energy losses as well as operational costs in the supply chain, considerably increases the price of the electricity derived from this fuel. Transport efficiency, measured by the ship size, capacity, and speed, has a moderate impact on both fossil fuels. Considering the moderate sensitivity to the carbon tax for coal, and the low sensitivity to NG, it can be concluded that its application would have only a moderate impact unless the taxation was drastically increased. The effect of the carbon tax on the competitiveness of green ECs is analyzed in more detail in Sec. \ref{sec:Break-even Analysis}.

Opposite to fossil fuels, which have to be extracted, the zero-carbon ECs, \ce{H2} and iron, have to be produced from electricity, whose price is clearly indicated as the second and third in the sensitivity of Fig.~\ref{fig:sensitivity}. It shows the reliance on access to cheap renewable energy, i.e.~as found in sunny arid regions (photovoltaic potential) or windy coastal areas (wind energy potential), which can justify long distance transport of the ECs. The efficiency of the production process is notably also highly influential. Unsurprisingly, sensitivities associated with plant equipment and infrastructure, such as CC and CRF, show a significant overall impact on the LCOE. It is interesting to notice, though, that the efficiencies of the production processes of green \ce{H2} and green \ce{Fe} only show approximately half the impact on the cost compared to the utilization efficiency. This is due to the high losses in efficiency associated with the utilization step and indicates that investing in research and development for efficient technologies for the thermochemical conversion of \ce{H2} (such as gas turbines) and iron (such as retrofit coal-fired power plants) is key for the economic competitiveness of the respective EC. It should be noted that the LCOE for the \ce{H2} supply chain shows the highest sensitivity to the ship's speed and size, which is in line with the discussion in Sec.~\ref{sec:maritimeTransport}. Together with the scaling ratio V$_{\ce{H2}}$/V$_{LNG}$, which compares the cargo volume of \ce{H2} ships to LNG ships of the same outer dimensions, the ship size directly influences how much EC can be transported per trip, while the speed determines the supply chain time. On the other hand, the green iron supply chain shows a lower sensitivity to transport parameters, indicating that its transport, even with minimal modifications--i.e. the type of fuel--, will unlikely become a barrier for the technology.

\FloatBarrier

\subsubsection{Uncertainty analysis} \label{sec:uncertainty_analysis}

Besides the sensitivity, the uncertainties associated with the variables used for the overall supply chain assessments can influence the results. Therefore, an attempt is being made to quantify the influence of assumptions, particularly for potential future energy supply chains, namely the supply chains for green \ce{H2} and \ce{Fe}. Predictions about future technological developments are challenging in an interconnected and globalised world. However, this uncertainty analysis can still yield insights into decisive aspects to be analysed and monitored when implementing such clean energy supply chains.

\begin{figure}[h]
\begin{center}
\includegraphics[width=0.5\textwidth]{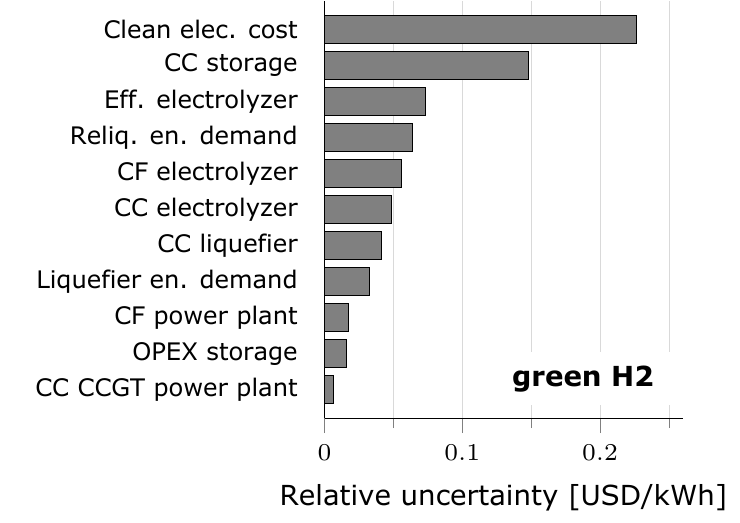}%
\includegraphics[width=0.5\textwidth]{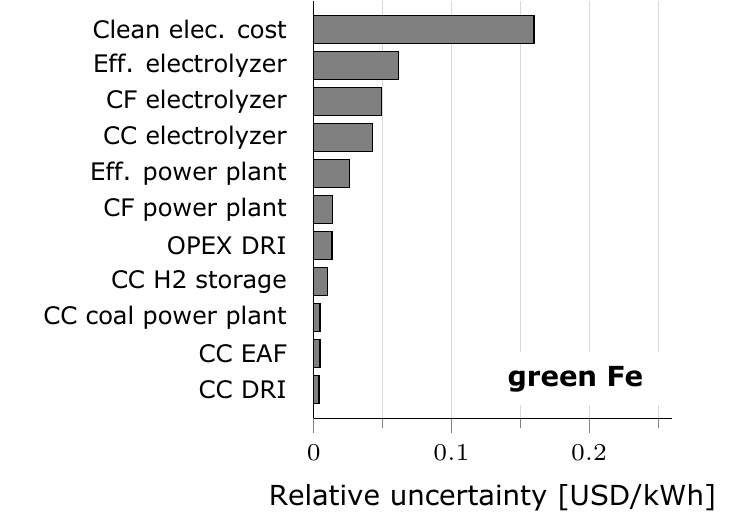}%
\end{center}
\vspace{-6pt}
\caption{Uncertainty analysis of LCOE for green \ce{H2} and green iron.}
\label{fig:uncertainty}
\end{figure}

Relative uncertainties of the LCOE are evaluated as described in Sec.~\ref{sec:uncertainty_analysis_theory} and the results for the supply chains for green \ce{H2} and green \ce{Fe} are shown in Fig.~\ref{fig:uncertainty}. Note that, according to Eq.~(\ref{eq:reluncertainty}), variables which exhibit a significant impact on the supply chain (high sensitivity in Fig.~\ref{fig:sensitivity}) but only show a small uncertainty induce only a small relative uncertainty on the LCOE and thereby affect the ranking of the parameters. 

Inspecting Fig.~\ref{fig:uncertainty}, the largest uncertainty for the LCOE of green \ce{H2} and iron is, by far, caused by the clean electricity costs. Both supply chains are sensitive to this parameter and show high uncertainty. While the global economy relies on fossil fuels and no relevant large-scale production facilities exist for zero-carbon ECs, mostly demonstration plants and theoretical assessments exist for clean electricity costs for potential clean energy exporters. Part of the high uncertainty shown in Fig.~\ref{fig:uncertainty} also includes variability in clean energy production potential, which stems from geographic location, climate, and local infrastructure, which is also reflected in the relatively high uncertainty of the capacity factor for the electrolyzer. These high relative uncertainties underline the importance of suitable locations for the production of zero-carbon ECs, and it can also be understood as a great opportunity if producers succeed in implementing very efficient \ce{H2} production.

The second largest uncertainty for green \ce{H2} relates to the capital costs of the storage. Interestingly, this uncertainty also affects the iron supply chain since it is assumed that the iron transporting vessels would use green hydrogen as fuel, which depends on the production and storage steps in the hydrogen supply chain.
Additionally, the efficiency of the electrolyzer for hydrogen production, which is part of both EC's supply chains, induces significant uncertainty. While the electrolyzer technology is currently scaled to large dimensions and will conceivably improve in the coming years, this uncertainty will soon become smaller. Relevant differences between the uncertainty assessments for the LCOE of the supply chains of green \ce{H2} and iron are \ce{H2}-(re)liquefaction and capital costs. \ce{H2}-liquefaction does not play a role for green iron, but this process is important for the green \ce{H2} supply chain. Efficient and cost-effective liquefaction still has to be implemented at scale and, if improved, could yield significant benefits for the LCOE of \ce{H2} in general. On the other hand, the capital costs for the hydrogen supply chain lead to further uncertainties since most of the infrastructure and equipment (production facilities, gas turbines, ships) still have to be developed and later constructed. Given that coal power plants can be retrofitted for iron and that suitable ships already exist, CCs lead to notably less uncertainty for this EC in the present model.

\FloatBarrier

\subsubsection{Break-even analysis} \label{sec:Break-even Analysis}
\begin{figure}[htb]
\begin{center}
\includegraphics[width=0.95\textwidth]{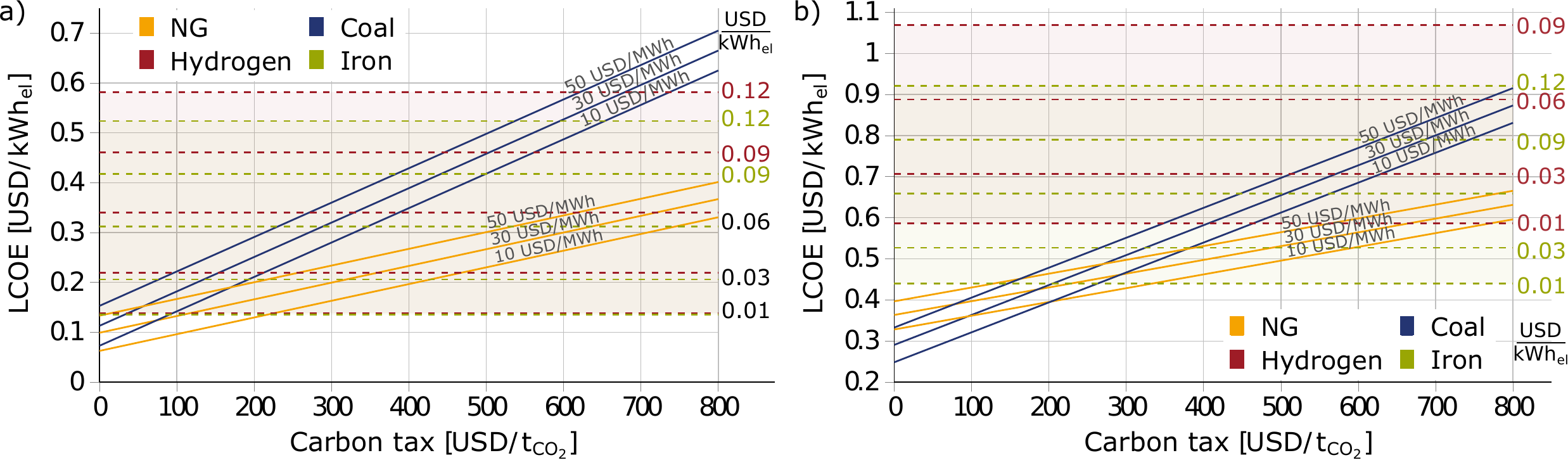}\quad%
\end{center}

\vspace{-6pt}
\caption{Break-even analysis assuming a transport distance of roughly 8000km (YAN-HAM). LCOE of coal and NG supply chains are shown as function of the carbon tax for different fuel prices. The LCOE of iron and hydrogen are shown as variations of renewable energy costs (horizontal lines). a) best-cases for all ECs, b) averaged-cases for all ECs.}
\label{fig:Break Even analysis}
\end{figure}

The previous analyses indicate that the cost of producing renewable energy significantly impacts the economic performance of carbon-free ECs. On the other hand, the LCOE for conventional EC supply chains is moderately (i.~e.~NG) to strongly (i.~e.~coal) influenced by the cost of their raw materials and the carbon taxation policy aimed at reducing \ce{CO2} emissions. To identify the cost combinations at which green EC have lower LCOE than conventional EC supply chains, a break-even analysis is performed for both the best and average scenarios, as illustrated in Fig.~\ref{fig:Break Even analysis}. 

Both the figures show that coal is more affected by carbon pricing than NG due to its higher carbon intensity and that the hydrogen supply chain is more sensitive to the price of renewable energy than iron due to its higher share of energy costs in the LCOE.
In the best case scenarios (Fig.~\ref{fig:Break Even analysis} a)) iron and hydrogen can already be competitive with LCOE of approximately 0.14\,$\mathrm{USD} \cdot \mathrm{kWh_{el}}^{-1}$ at very low prices for renewable electricity (e.g. 0.01\,$ \mathrm{USD} \cdot \mathrm{kWh_{el}}^{-1}$) and low carbon pricing of 90\,$ \mathrm{USD} \cdot \mathrm{t}_{\ce{CO2}}^{-1}$ with coal at a fuel price of 10$\,\mathrm{USD} \cdot \mathrm{MWh}^{-1}$ and the corresponding assumptions (c.f.~\ref{sec:assumptions}). For NG, the intersection is roughly at 220\,$\mathrm{USD} \cdot \mathrm{t}_{\ce{CO2}}^{-1}$. At NG prices of 50\,$\mathrm{USD} \cdot \mathrm{MWh}^{-1}$ (which is in the order of recent prices of the Dutch TTF Natural Gas Future, see Table~\ref{table:FF_production}) and a carbon tax of 220\,$\mathrm{USD} \cdot \mathrm{t}_{\ce{CO2}}^{-1}$ the supply chains for iron and hydrogen become competitive with NG at a price around 0.03\,$\mathrm{USD} \cdot \mathrm{kWh_{el}}^{-1}$ for renewable energy.

In comparison to the best-case assumptions, which include the lowest capital and operational expenses, highest efficiencies, and high capacity factors for the corresponding power plants, the averaged results between the best- and worst-case scenarios (Fig.~\ref{fig:Break Even analysis} b) show a significant higher offset (for all ECs) and impact of the costs for renewable energy, especially for hydrogen. The higher uncertainty with hydrogen-related parameters results in a significantly higher LCOE for hydrogen compared to iron.
In this scenario, hydrogen can only be competitive with NG at extraordinary carbon taxes and at very high carbon taxes (e.g. 400\,$\mathrm{USD} \cdot \mathrm{t}_{\ce{CO2}}^{-1}$ with coal. On the other hand, the supply chain of iron can get competitive at carbon taxation around (250\,$\mathrm{USD} \cdot \mathrm{t}_{\ce{CO2}}^{-1}$ with coal and NG.

The break-even analysis illustrates that carbon-neutral ECs such as green hydrogen and iron can become competitive with conventional ECs at low costs of renewable energy. However, it is unlikely that these green ECs will reach this level of competitiveness without implementing a carbon taxation system to account for some of the environmental impact by conventional ECs. The European Union Emission Trading System serves as an example of a carbon pricing system (EU carbon prices (end 2022) are in the order of 90\,EUR $\cdot \ \mathrm{t}_{\ce{CO2}}^{-1}$ ~\cite{Ember.2022}), with recent announcements of tighter \ce{CO2} allowances~\cite{EU.2022c} leading to higher carbon prices and consequently increased competitiveness for alternative green ECs.

\FloatBarrier
\section{Conclusions}
\label{sec:Conclusions}

In the present work, a techno-economic assessment for the long-distance energy supply chains for carbon-free electricity generation of the two promising zero-carbon energy carriers (ECs), \ce{H2} and iron, is presented. In particular, the thermodynamic efficiency (energetic assessment), \ce{CO2} emissions (environmental assessment), and the levelized cost of electricity (LCOE, economic assessment) are analyzed. For comparison, analogous assessments are being carried out for the supply chains of the well-established fossil fuels, coal and natural gas, for which international long-distance trade has been established for decades. 

\noindent With respect to the zero-carbon energy carriers, the following conclusions can be drawn:
\begin{enumerate}
    \item \textbf{Energetic efficiencies:} Green iron exhibits energy efficiencies that are comparable to those of green hydrogen. Optimizing all processes along the energy supply chains and depending on transport distances, the potential for the power-to-power energy efficiency is 19\%--29\% for iron and 16\%--32\% for hydrogen.
    \item \textbf{Influence of transport and storage:} For short-distance transport and immediate usage, hydrogen might exhibit a competitive advantage since iron requires additional processing steps to be regenerated from its oxides. For long-distance transport and/or long storage times, however, these advantages become overcompensated by the more favorable transport and storage characteristics of iron compared to hydrogen.
    \item \textbf{Economic assessment:} Our analysis indicates the advantages of iron compared to hydrogen, mainly due to the retrofitting potential of existing infrastructure and more favorable transport and storage characteristics. This comparison is, however, subject to many uncertainties, which largely stem from the costs associated with the required large-scale infrastructure for iron (iron oxide regeneration plants) and for hydrogen (liquefiers, storage, transport) that does not exist today.
    \item \textbf{Cost comparison to fossil fuels:} Realizing the techno-economic potentials along the considered zero-carbon energy supply chains, it is conceivable that iron and hydrogen can become cost-competitive with fossil fuels. However, this largely depends on low renewable energy prices and the introduction of carbon taxation. The analysis further shows that access to cheap renewable energy can overcompensate the costs associated with long-distance transport.
\end{enumerate}

The current study considers overarching characteristics of energy supply chains and does not account for specifics on the system scale of individual plants, localized infrastructure for short-range transport, or climate/weather effects, some of which are valuable aspects for follow-up work. As the technologies of zero-carbon energy carriers evolve, future research could further be carried out along the following directions:
\begin{itemize}
    \item \textbf{Regeneration of iron:} The reduction of iron is one of the challenges for realising the metal energy cycle and is subject to research and development. It can be assumed that synergies with efforts to decarbonise steel production will soon lead to technological progress for this process. Incorporating operating data for clean reduction plants will reduce uncertainties in the assessment of iron as an energy carrier.
    \item \textbf{Implementation times:} Hydrogen supply chains include processes which have yet to be realised at large scales (i.e.~liquefaction, transport, and storage), which raises questions about the implementation times for the required infrastructure. On the other hand, iron could largely profit from retrofitting infrastructure formerly used for coal. Opportunities and challenges for green \ce{H2} and green iron are linked to sub-processes with very different technology readiness levels (TRL) along their respective supply chains, influencing the implementation times. It remains to be investigated how this might affect the adoption of these zero-carbon energy carriers, given that decarbonisation goals for many industries require effective measures sooner than later.
    \item \textbf{In-depth economic modeling of intermittent renewable energy supply:} The economics of variable renewable energy (VRE) plants, such as the recycling facilities for iron oxides, are more complex than the current analysis might suggest. For instance, it might be beneficial in specific scenarios to sell energy from renewable sources directly into the electricity market at the plant's location rather than storing it in a chemical energy carrier and then shipping it to other countries. An investigation of such energy markets' drivers requires sophisticated economic modeling considering, for instance, spot market pricing or long-term contracts. Such investigations could help refine and improve LCOE estimates for emerging zero-carbon energy carriers such as hydrogen and iron.
\end{itemize}

Considering the substantial funding programs for green hydrogen technologies from governments worldwide, it is safe to assume that a \ce{H2}-infrastructure will be implemented in many countries. The present study further suggests that also green iron shows the potential to become an important energy carrier for long-distance trade in a globalized clean energy market and should therefore be discussed as a technological option.

\section*{Acknowledgments}
This work was funded by the Hessian Ministry of Higher Education, Research, Science and the Arts - Clean Circles cluster project. The authors thank Prof. Andreas Dreizler for his advice and valuable feedback. The authors further acknowledge Sunil Chapagain, Evrim Cicek, Lukas Schleidt, Bipal Shrestha, Kieu-Ly Tran and Ying Lin for support with the data collection and literature review.

\clearpage

\section*{\textbf{Appendix: Detailed formulation}}

\renewcommand{\thesection}{A\arabic{section}}
\renewcommand{\thefigure}{A\arabic{figure}}
\renewcommand{\thetable}{A\arabic{table}}
\renewcommand{\theequation}{A\arabic{equation}}

\setcounter{section}{1}
\setcounter{subsection}{0}
\setcounter{figure}{0}
\setcounter{table}{0}
\setcounter{equation}{0}

For a certain vessel with its cargo measured in volume or mass, the energy carrier (EC) mass is given by:
\begin{equation}    \label{eq:MECfromCargo}
   m_{EC} = m_{cargo} = \rho_{EC} \cdot V_{cargo} \ [\mathrm{kg}]
\end{equation}

If boil off gas (BOG) occurs at any phase of the transport, i.e. during the long-range transport and the intermediate storage, the mass of the EC at a certain phase can be given as a function of the initial cargo mass,  multiplied by the boil off rate (BOR) of every stage between that phase and the ship loading. For the mass before and after the ship loading, equations~\ref{eq:MECbeforeShip} and~\ref{eq:MECafterShip} can be used, respectively:
\begin{equation}    \label{eq:MECbeforeShip}
   m_{EC} = m_{cargo} \cdot \Pi{(1 + BOR_i)^{t_i}} \ [\mathrm{kg}]
\end{equation}
\begin{equation}    \label{eq:MECafterShip}
   m_{EC} = m_{cargo} \cdot \Pi{(1 - BOR_i)^{t_i}} \ [\mathrm{kg}]
\end{equation}

The primary energy content for a given mass of the EC is:
\begin{equation}    \label{eq:PEfromCargo}
   PE_{EC} = m_{EC} \cdot LHV_{EC} \ [\mathrm{kWh}]
\end{equation}

\section{\textbf{Production}}
\label{sec:production}
To produce a synthetic EC, a certain amount of energy input is required, consisting of the internal energy of the feedstock and the energy needed to synthesize it. This energy input can be calculated by determining the overall energetic efficiency of the production process.
\begin{equation}    \label{eq:WprodfromPE}
   W_{prod} = \frac{PE_{EC}}{\eta_{prod}} \ [\mathrm{kWh}]
\end{equation}

Such syntheses process can have a \ce{CO2} intensity, the ratio between the mass of the emission and the mass of EC generated. In the present work, such value is taken as a constant, based on the report from the IEA~\cite{thefutureofhydrogen}. Therefore, the \ce{CO2} emitted per ship cargo can be measured as:
\begin{equation}    \label{eq:CO2prodfromMEC}
   E_{\mathrm{\ce{CO2}}_{prod}} = I_{\mathrm{\ce{CO2}}{_{prod}}} \cdot m_{EC} \ [\mathrm{kg}_{\mathrm{\ce{CO2}}}]
\end{equation}

\section{\textbf{Liquefaction}}

Liquefaction processes can be assumed as demanding a constant energy per unit of mass:
\begin{equation}    \label{eq:W_Liq}
   W_{liq} = m_{EC} \cdot w_{liq} \ [\mathrm{kWh}]
\end{equation}

This energy is considered to be provided solely by electricity. In this sense, the process will have associated \ce{CO2} emissions solely related to the emission intensity of the electricity generation:
\begin{equation}    \label{eq:CO2_liq}
   E_{\mathrm{\ce{CO2}}_{liq}} = I_{\mathrm{\ce{CO2}}{_{elec}}} \cdot W_{liq} \ [\mathrm{kg}_{\ce{CO2}}]
\end{equation}

\section{\textbf{Storage}}

To temporarily store ECs in ports, before their export by ship or their distribution locally, a certain amount of energy is necessary to maintain the substances in transportable form (i.e. re-liquefaction of the BOG for NG and \ce{H2}.). In the present work, the required time of loading and unloading, from the pipelines or trucks to the export terminal facilities and later to the ship, and from the ship to the import terminal facilities and later to the pipelines or trucks for the delivery, is estimated to be 6 days. Considering the combined energy to be provided solely by electricity, depending on the mass and the time, the necessary energy for the storage can be given by:
\begin{equation}    \label{eq:W_storefromMEC}
   W_{store} = m_{EC} \cdot \dot w_{store} \cdot \mathrm{t}_{store}\ [\mathrm{kWh}]
\end{equation}

If the only energy source is electricity, the \ce{CO2} emissions are given by the emission's intensity of the power generation:
\begin{equation}    \label{eq:CO2_store}
   E_{\mathrm{\ce{CO2}}_{store}} = I_{\mathrm{\ce{CO2}}{_{elec}}} \cdot W_{store} \ [\mathrm{kg}_{\ce{CO2}}]
\end{equation}

The costs for storage are the sum of the CAPEX, the fixed OPEX, and the electricity costs. The CAPEX is given by the annualized specific costs considering the operational days per year. In contrast to the CF for the production, as the number of days of operation increases ($t_{store}$), so do the associated costs.  
\begin{equation}    \label{eq:CAPEX_store}
   CAPEX_{store} = CRF \cdot CC_{tank} \cdot \frac{t_{store}}{t_{year}} \ [\mathrm{USD} \cdot (\mathrm{kg}_{EC}\mathrm{year}^{-1})]
\end{equation}

\section{\textbf{Long-range transport}}
The long-range transport in ships, from the export terminal to the import terminal, is required for moving the EC between the locations. Excluded from the analysis are costs and energy for loading and unloading. Furthermore, ships that transport liquid fuels are considered to use thermal isolation in order to keep the conditions, being more energy-efficient but subject to boil-off, either losing cargo or using it to fuel their engines. The energy demand of such ships is given by the ship's engine power capacity, its efficiency or heat rate, and the travel time:
\begin{equation}    \label{eq:W_LD}
   W_{trans} = \frac{\dot{W}_{engine}}{\eta_{engine}} \cdot 24 \frac{{h}}{{day}} \cdot {t}_{trans} \ [\mathrm{kWh}] 
\end{equation}

If the travel time is taken, considering a certain constant speed, and that the total travel distance is twice the distance between the ports, therefore considering the round-trip, it can be calculated by the equation below:
\begin{equation}    \label{eq:t_LD}
   \mathrm{t}_{trans} = 2 \cdot \frac{d_{trans}}{24 \cdot v_{ship}} \ [\mathrm{days}]
\end{equation}

The \ce{CO2} emissions of each trip depend on the emission intensity of the fuel consumed. The total \ce{CO2} can be given by:
\begin{equation}    \label{eq:CO2_LD}
   E_{\mathrm{\ce{CO2}}_{trans}} =  W_{trans} \cdot I_{\mathrm{\ce{CO2}}{_{fuel}}}\ [\mathrm{kg}_{\ce{CO2}}]
\end{equation}

In the case of liquefied ECs, the net mass of EC after the trip, is given by:
\begin{equation}   \label{eq:MECafterShip2}
   m_{EC_{trans}} = m_{cargo} \cdot (1 - BOG_{trans})^{t_{trans}/2} \ [\mathrm{kg}]
\end{equation}

In the case of ships transporting and consuming the EC (i.e. \ce{H2} or \ce{NG}), the boil-off value is the greater value between the natural boil-off and the forced boil-off due to the engine consumption:
\begin{equation}    \label{eq:BOG_H2}
   BOG_{trans} = max(BOG_{ship}, BOG_{forced})\ [\mathrm{wt. \%} \cdot {day}^{-1}]
\end{equation}

In those cases, the energy of the extracted cargo is considered equal to the energy demand of the ship:
\begin{multline}    \label{eq:BOG_f0}
   W_{trans} = (m_{cargo} - m_{EC_{trans}}) \cdot LHV_{EC} = \\ = (m_{cargo} - m_{cargo} \cdot (1 - \frac{BOG_{forced}}{100})^{t_{trans}/2}) \cdot LHV_{EC}\ [\mathrm{kWh}]
\end{multline}

Therefore, the forced boil-off can be given by:
\begin{equation}    \label{eq:BOG_f1}
   BOG_{forced} = 100 \cdot \left(1 - \left(1 - \frac{W_{trans}}{m_{cargo} \cdot LHV_{EC}}\right)^{2/t_{trans}}\right)\ [\mathrm{\%}]
\end{equation}

If the natural boil-off in energy content is equal or greater than the energy consumed, or if the only fuel is the EC of the cargo, the shipping costs are given by:
\begin{equation}    \label{eq:C_LD2}
   C_{trans} = CAPEX_{trans} + OPEX_{trans} + 2 \cdot C_{canal} \ [\mathrm{USD}]
\end{equation}

\subsection{Capital costs of ships} \label{sec:CC_ships}
To define the capital costs of the ships used in the present work, the formulation from the work of Mulligan \cite{mulligan2008} was used. In this formulation, taken from the fitting of a pool of vessel sizes and costs, uses an offset for each ship, the producer price index (PPI) for ship building and the deadweight tonnage of the vessel (DWT) in $10^6$\,kg.
\begin{multline}    \label{eq:CC_ship}
   CC_{ship} = offset_{ship} + 2.6 \cdot PPI + 1.8055 \cdot DWT - 0.01009 \cdot DWT^2 + \\ + 0.0000189 \cdot DWT^3 \ [10^6 \ \mathrm{USD}]
\end{multline}

For the L\ce{H2} ship, considering that such ships are still not available in commercial scale, it is possible to calculate the capital cost as a function of the cost of a similarly-sized LNG ship, assuming a multiplier:
\begin{equation}    \label{eq:CC_LH2}
   CC_{L\ce{H2}} = M_{L\ce{H2}/LNG} \cdot CC_{LNG} \ [10^6 \ \mathrm{USD}]
\end{equation}

The offset for each ship can be given by Table \ref{table:offsets}.

\begin{table}[H]\small
\caption{Offset for equation \ref{eq:CC_ship}, by ship type \cite{mulligan2008}}
\centering
\begin{tabular}{l  c c c  c c}
    \hline
    EC & Unit & Natural Gas & Coal & Iron \\
    \hline
    Ship type & [-] & LNG ship & dry bulk carrier & dry bulk carrier \\
    Offset & [$10^6$ USD] & -253.012 & -418.202 & -418.202 \\
    \hline
\end{tabular}
\label{table:offsets}
\end{table}

\section{\textbf{Electricity generation}}

When utilized for electricity generation in a power plant, the net output depends on the net efficiency of the corresponding power plant.
\begin{equation}   \label{eq:W_elecfromReconv}
  W_{elec} = \eta_{elec} \cdot m_{EC} \cdot LHV_{EC} 
  \ [\mathrm{kWh_{el}}]
\end{equation}

The direct \ce{CO2} emissions of this operation depend on the emission intensity of the EC:
\begin{equation}   \label{eq:CO2_elec}
  E_{\mathrm{\ce{CO2}}_{elec}} =  I_{\mathrm{\ce{CO2}}{_{EC}}} \cdot \cdot m_{EC} \ [\mathrm{kg}_{\ce{CO2}}]
\end{equation}

\FloatBarrier 

\clearpage

\footnotesize{
 \baselineskip 9pt
\bibliographystyle{unsrtnat}
\bibliography{literature.bib}}

\clearpage

\end{document}